\documentclass{sig-alternate}

\DeclareMathOperator{\dom}{Dom}

\newcommand{\eop}{\hfill $\Box$}

\newcommand{\ccc}{{\cal C}}
\newcommand{\barccc}{\bar{\cal C}}
\newcommand{\ccs}{{\cal S}}
\newcommand{\barccs}{\bar{\cal S}}
\newcommand{\restrictedto}{\upharpoonright}
\newcommand{\con}{_{con}}
\newcommand{\non}{_{non}}

\newtheorem{numer}{\hspace{-1mm}($\spadesuit$\hspace{-1mm}}
\newcommand{\piki}[2]{\vspace{-1mm}
\begin{numer}
\hspace{-1.5mm}{\em )}
\label{#1} 
#2 
\end{numer}
\vspace{-3mm}
}

\newcommand{\rpiki}[1]{($\spadesuit$ \hspace{-1.5mm} \ref{#1}.)}

\begin{document}

\clubpenalty=10000
\widowpenalty = 10000

\title{On the BDD/FC Conjecture \titlenote{Supported by Polish Ministry of Science and Higher Education NCN grant N N206 371339.}}

\numberofauthors{2} 
\author{
%
%
\alignauthor
Tomasz Gogacz\\
       \affaddr{Institute of Computer Science}\\
       \affaddr{University of Wroclaw}\\
       \affaddr{Poland}\\
       \email{gogo@cs.uni.wroc.pl}
\alignauthor
Jerzy Marcinkowski\\
       \affaddr{Institute of Computer Science}\\
       \affaddr{University of Wroclaw}\\
       \affaddr{Poland}\\
       \email{jma@cs.uni.wroc.pl}
}

\date{30 July 1999}

\newdef{definition}{Definition}
\newdef{example}{Example}
\newdef{remark}{Remark}
\newtheorem{conjecture}{Conjecture}
\newtheorem{theorem}{Theorem}
\newtheorem{lemma}{Lemma}

\begin{abstract}
Bounded Derivation Depth property (BDD) and Finite Controllability (FC) are two 
properties of sets of datalog rules and tuple generating dependencies (known as Datalog$^\exists$ programs), 
which recently attracted  some attention. 
We conjecture that the first of these properties implies the second, and support this conjecture 
by some evidence proving, among other results, that it holds true for all theories over binary signature.
\end{abstract}

{Mathematical Logic}
 {rule-based databases, query processing}

\section{Introduction}
 Tuple generating dependencies (TGDs), recently known also as Datalog$^\exists$ rules, are studied in
 various areas, from database theory to description logics, and in various contexts. The context we are 
 interested in here, is  computing certain answers 
 to queries in the situation when  some semantical information about the database is known, and  represented by some theory $\cal T$ (or a Datalog$^\exists$  program), 
consisting of existential TGDs and plain datalog rules, 
but it is assumed that our knowledge of the database facts  is incomplete 
(this  is known as the open-world assumption).

In this paradigm, for a database instance $D$ (understood here as a set of facts -- atomic formulas),
the semantics of $D$, in presence of $\cal T$ is defined as the 
(*) set of all the database instances $\bar D$ which are supersets 
of $D$ and satisfy $\cal T$. In other words,  we are  interested  whether, for a given query\footnote{Whenever we say 
"query" in this paper we mean a conjunctive query without negation. Whenever we say ''TGD'' we mean a single-head tuple generating dependency.}  $\Phi$, 
it holds that  ${\cal T}, D\models \Phi$. 

The problem is of course undecidable in general, so some restricted classes of theories are being studied.  For example 
Linear Datalog$^\exists$ programs, which consist of TGDs which, as the body, have a single atomic formula, were studied in \cite{R06},
Guarded Datalog$^\exists$, being an extension of Linear (see Section \ref{guarded} for more details) was analyzed in \cite{BGO10}
and  Sticky Datalog$^\exists$ programs were introduced (in two flavors) in \cite{CGP10} and \cite{CGP10'}.

As it turns out, decidability of query answering is not that hard to prove for theories from these classes.
But there are good reasons why we would like to have more than just decidability. The desired properties of $\cal T$ are 
(among others) Bounded Derivation Depth property (BDD) and Finite Controllability (FC).

The theory $\cal T$ has the Finite Controllability property (for short: "$\cal T$ is FC"), 
if the expression "the set of all database instances" in the definition (*) above can be equivalently replaced by, 
more natural from the database point of view, "the set of all finite database instances".
To be more precise:

\begin{definition} $\cal T$ is FC if for each database instance $D$ and each query $\Phi$, if  ${\cal T}, D\not\models \Phi$ then
there exists a database instance $\cal M$ such that ${\cal M}\models D,\cal T$ but ${\cal M}\not\models \Phi$.
\end{definition}
The difficult technical results in \cite{R06} (solving an old problem stated in \cite{JK84}) 
and in \cite{BGO10} concern Finite Controllability of, respectively,  Linear and Guarded Datalog$^\exists$.
(Actually, the result in \cite{R06} is stated in terms of Inclusion Dependencies rather than TGDs, 
which, in this context, is another language to talk about the same thing.)
The question if the  Sticky Datalog$^\exists$ programs are FC was left as an open problem in \cite{CGP10} and was solved, positively, 
in \cite{GM13}. 

The theory $\cal T$ has the  Bounded Derivation Depth property (or just: "$\cal T$ is BDD") if it admits positive first order query rewriting.
In other words:

\begin{definition}\label{bedede}
$\cal T$  is BDD if for each query $\Phi$ there exists a union of conjunctive queries $\Phi'$ such that
for every database instance $D$ the equivalence :  ${\cal T}, D\models \Phi$ if and only if $D\models \Phi'$ holds.
(This is the definition we actually need here, but not the standard one. See Section \ref{preli} for an equivalent, more standard definition.
)
\end{definition}

 This means that 
instead of  computing the 
answer to $\Phi$ over the infinite set of databases having $D$ as their subset (or instead of computing 
the answer to $\Phi$ over the infinite database $Chase(D, {\cal T})$ -- see Section \ref{preli})  it is enough to compute the answer to $\Phi'$ over the known finite database $D$. 
 There is no need to explain how  desirable in the database context BDD is, so many of the good classes of Datalog$^\exists$ programs 
(including Linear  Datalog$^\exists$ and Sticky Datalog$^\exists$) are tailored
to have this property. It is worth mentioning that, while of course BDD is an undecidable property of $\cal T$, still in all 
practical situations we know about, proving the statement  
"all the programs from class $\cal C$ are BDD" is an easy exercise (if it is true). This is 
in sharp contrast to Finite Controllability which is, as we mentioned above, typically quite hard to prove.

BDD is typically easy to prove. FC is hard to prove. But each time we had a class of BDD theories, finally we were
able to show that this class is also FC. This leads to a conjecture we would like to state here:

\begin{conjecture}[The BDD/FC conjecture]\label{hipo}~\\
If some theory $\cal T$, being a set of existential TGDs and plain datalog rules, is BDD then it is also FC.
\end{conjecture}

The evidence we support our conjecture with is: 

\begin{theorem}[The main  result of this paper]\label{glownewintro}~\\
Conjecture \ref{hipo} is true for programs over binary signature.
\end{theorem}

The proof of Theorem \ref{glownewintro} is the main technical contribution of this paper and is,
as we believe, quite difficult. It is 
presented in Section \ref{dowod} but relies
on a system of tools developed in Sections \ref{teoria} and \ref{odlozona}.

Finally, in Section \ref{dyskusja} we discuss the possible applications of our tools and their limitations,
In subsection \ref{ubranie} we show that Theorem \ref{glownewintro} can be extended also to 
quite a wide class of non-binary theories
(see Theorem \ref{niebinarne}). In subsection  \ref{why} we explain however, why our techniques 
do not seem to extend to the proof of 
Conjecture \ref{hipo} in general. 
In subsection \ref{guarded} we show how Guarded Datalog$^\exists$ programs can be seen as binary programs, 
and how our techniques can be easily applied in this context.

\subsection{TGDs and Chase -- preliminaries}\label{preli}

Let us remind the reader that  a TGD is a formula of the form $\forall \bar x \; (\Phi(\bar x) \Rightarrow \exists y \; Q(y,\bar y))$ where 
$\Phi $ is a conjunctive query, $Q$ is a relation symbol, $\bar x, \bar y$ are tuples of variables and
$\bar y\subseteq \bar x$ (see Section \ref{multihead} for a comment on the multi-head TGDs). The universal quantifier in front of the formula is usually omitted.

Finite sets consisting of existential TGDs and plain datalog rules will be called theories. 
 

For a theory $\cal T$ and a database instance $D$ we denote by  $Chase^1(D, {\cal T})$  the result of the following operation.
For each tuple $\bar x$ in $D$ satisfying a body of an rule $t_i=\forall \bar x \; (\Phi(\bar x) \Rightarrow \exists y \; Q(y,\bar y))$ form $\cal T$, such that there is no $y \in D$ satisfying $D\models Q(y,\bar y))$,
we simultaneously add new constant $c_{t_i,\bar x}$ into database and an atom $Q(c_{t_i,\bar x}, \bar y)$.

Then define $Chase^{i+1}(D,\cal T)$ as $Chase^1(Chase^i(D, \cal T), \cal T)$ and by Chase denote $\bigcup_i Chase^i(D,\cal T)$, which is the least fixpoint of the $Chase^1$ operator.

Clearly, we have $Chase(D, {\cal T})\models D, \cal T$, but there is no reason to think that $Chase^i(D, {\cal T})\models \cal T$
for any $i\in \mathbb N$. Note that the chase we consider in this paper is  the non-oblivious one -- new elements are only 
created if needed, as opposed to the blind Chase, which creates a new witness each time it is demanded.

Since $Chase(D, {\cal T})$ is a ''free structure'',  it is very easy to see that
 for any query $\Phi$ 
 (being a UCQ -- a union of positive conjunctive queries) 
  $D,{\cal T}\models \Phi$ (which reads as ''$\Phi$ is certainly true in $D$, in presence of $\cal T$''),
if and only if $Chase(D, {\cal T})\models \Phi$.

 A set $\cal T$ of TGDs is  usually said to have 
 Bounded Derivation Depth property if for each query $\Psi$, there is a constant $k_\Psi\in \mathbb N$, such that for each database instance  $D$
 if $Chase(D, {\cal T})\models \Psi$ then $Chase^{k_\Psi}(D, {\cal T})\models \Psi$.
It is easy to see (\cite{CGT09}) that  this definition of the BDD property 
is equivalent to Definition \ref{bedede}.

{\bf Notations.} When we say that $\ccc $ is a structure we may mean both, the set of elements and the set of atoms of $\ccc$.
If we feel this may cause confusion we write $\dom(\ccc)$ for the set of elements of $\ccc$. 
By $\ccc\models R$ (or $\ccc\models \psi$)  we mean that an atom $R$ (or a formula $\psi$)
 is true in $\ccc$. By $\ccc_1\models \ccc_2$ we mean that each atom of $\ccc_2$ is an atom of $\ccc_1$. 
For a structure $\ccc$ and a set $A$ (or a signature $\Sigma$)
by $\ccc \restrictedto A$ (resp. by $\ccc \restrictedto \Sigma $) we mean the structure consisting of such atoms 
$R(\bar a)$ that  $\ccc \models R(\bar a)$ 
and $\bar a\subseteq A$ (resp. $R \in \Sigma$). 
For a structure $\ccc$ over some signature $\Sigma$ by $\ccc\con $ we mean, depending on context,
 the set of  elements of $\ccc$ which are
interpretations of constants from $\Sigma$ or the structure $\ccc\restrictedto \ccc\con$. Similarly, 
by $\ccc\non $ we mean
 the set of  elements of $\ccc$ which are
not incarnations of constants from $\Sigma$ or the structure $\ccc\restrictedto \ccc\non$.
  
In paper we consider only Boolean conjunctive queries. Sometimes free variables are omitted to keep the notation light.
In such cases one should treat them as existentially quantified.
(For example, for a query $\Phi(\bar x)$ the term $M \models \Phi$ should be read as $M \models \exists \bar x \; \Phi(\bar x)$).

\section{Types and projections}\label{teoria}

\subsection{The main ideas and\\ the structure of the proof}

In order to prove Theorem \ref{glownewintro} we need to construct, for a given BDD theory ${\cal T}$ over a binary signature,
 for a conjunctive
query $Q(\bar x)$ and for a finite structure $D$, such that $Chase(D,{\cal T})\not\models \exists \bar x\; Q(\bar x)$,
a new finite structure $M$,
such that $M\models D,{\cal T}$ but $M\not\models \exists \bar x\; Q(\bar x)$.

Such $M$ will always contain a substructure $M'$ being a homomorphic image of $Chase(D,{\cal T})$.
This $M'$ is easy to construct inside $M$ (if we had $M$): start from $D$, which is a substructure of both $M$
and  $Chase(D,{\cal T})$, and then mimic, inside $M$, all the applications of rules that led to the construction
of $Chase(D,{\cal T})$.

Isn't
$M'$ itself  always the finite model we are looking for? No, because it may very well happen that
by identifying elements of $Chase(D,{\cal T})$ the homomorphism (call it $q$) created new instances of the bodies
of the rules of $\cal T$ in $M'$, leading to the situation when applications of
rules is possible that were not  applied in $Chase(D,{\cal T})$.
For example suppose that $E(x,y),R(y,z)\Rightarrow U(y)$ is a rule of  $\cal T$ and
$Chase(D,{\cal T})\models E(a,b), R(b',c)$, but  $Chase(D,{\cal T})\not\models U(b)$ and $Chase(D,{\cal T})\not\models U(b')$.
Suppose also that $q(b)=q(b')$. Then the fact $U(q(b))$, which may not be homomorphic image of any fact
in $Chase(D,{\cal T})$,
is provable in $M$. This can lead to a process in which an answer to $Q$ is built in $M$, something we need to avoid.
This can also lead to infinite chase,
while we want $M$ to be a finite structure. Let us illustrate this problem with one more example:

\begin{example}\label{niepotrzebny}
Let $\cal T$ be a theory consisting of three rules:

$ E(x,y)\Rightarrow \exists z\; E(y,z) $

$E(x,y),E(y,z),E(z,x) \Rightarrow \exists t\; U(x,t) $

$ U(x,y)\Rightarrow \exists z\; U(y,z) $

and a database instance $D=\{E(a,b) \}$.

Then $Chase(D,{\cal T})$ is an infinite $E-chain$, beginning with $a$ and $b$. Consider $M'$ consisting
of elements $a,b$ and $c$ and atoms $E(a,b)$, $E(b,c)$ and $E(c,a)$. Then $M'$ is a homomorphic image of $Chase(D,{\cal T})$,
but is not itself a model of $\cal T$ -- the last rule, which was never triggered when $Chase(D,{\cal T})$
was built, can be used in $M'$.
 Moreover,  it is easy to see that $Chase(M',{\cal T})$ is an infinite structure.

\end{example}

The idea of the construction we present in this paper is to make sure that some sort of first order type of each
$a$ in $Chase(D,{\cal T})$ is
always the same as the type of its image $q(a)$ in the homomorphic image of $Chase(D,{\cal T})$. The definition of the
type should be tailored in such a way  that such preservation of types implies that
 no harmful new applications of rules
from $\cal T$  for $q(a)$ exist in $M$.

In Section \ref{teoria} we develop a sort of theory of positive types and their preservation. We built a framework in which
the Main Lemma (Lemma \ref{duzy}) can be expressed. In Section \ref{dowod} this Main Lemma is used to prove
Theorem \ref{glownewintro}. In Section \ref{odlozona} we prove the Main Lemma. Sections \ref{dowod} and \ref{odlozona} are independent
and can be read in any order.

{\bf The two most important technical tricks} of the paper can be found in proofs of Lemma \ref{wystarczy}
(in Section \ref{dowod}) and  Lemma \ref{normali}
(in Section \ref{odlozona}). In the proof of Lemma \ref{wystarczy}  we show how the assumption
that the theory $\cal T$ is BDD can be used. The trick in the proof of Lemma \ref{normali} relies on the construction,
presented already in Section \ref{born}, where we construct not just one finite structure,
but an infinite sequence of finite structures $M_n$ that in some sense converge to $Chase(D,{\cal T})$.
Then the idea is that if some query $\Psi$ is true in $M_{n+1}$ (which we do not like, as we do not want
too many queries to be true in the finite structures we construct)
then a query $\Phi$, being a  ''one-step normalized'' version of $\Psi$
may not be true in $M_{n+1}$ but it will be true in $M_{n}$.
This then implies that if $\Psi$ is true in all $M_n$ then its ''normal form'' also is. This ''converging to the Chase'' trick
is also used in our another paper \cite{GM13} and we believe it can have further applications.

\subsection{Positive types}

\begin{definition}\label{ptp-def}
Let $\cal C$ be a relational structure over  signature $\Theta$.
Let  $e\in \cal \ccc$ and  let $n$ be a natural number. We define
$ptp_n({\cal C}, e, \Theta)$ (which reads "positive n-type of $e$ in $\cal C$ over $\Theta$'') as the set of all
such conjunctive queries $\Psi(\bar x, y)$ that:
\begin{itemize}
\item
$|\bar x|<n$,
\item
 all  relations (and constants) used in
$\Psi$ are in $\Theta$
\item
${\cal C}\models  \Psi(\bar x, e)$.
\end{itemize}

We assume that equality belongs to each $\Theta$, which means that atoms of the form $x=c$ (but not of the form $x\neq c$),
where $x$ is a variable and $c$ is a constant from $\Theta$, are allowed in the queries.
\end{definition}

\begin{example}\label{drugi-niepotrzebny}
Let $\cal T$, $D$ and $M'$ be like in Example \ref{niepotrzebny}, and let $\Theta$ consist of $E$ and $U$.
 Then $ptp_2(Chase(D,{\cal T}) ), a, \Theta)$ equals $ptp_2(M', a, \Theta)$, and each of them consists of
the same two queries: $E(x,y)$ and $E(y,x)$. But  $ptp_3(Chase(D,{\cal T}) ), a, \Theta)$ does not equal $ptp_3(M', a, \Theta)$:~
the query $E(y,x_1) \wedge E(x_1,x_2)$ $\wedge E(x_2,y)$ belongs to the second of those two types but not to the first one.
\end{example}

\begin{remark}\label{pozna}
Notice that if $c$ is a constant\footnote{Strictly speaking, we mean a value of this constant in $\ccc$, but we
are not always going to make this distinction.}
 from $\Theta$, and if $a\neq c$ is any other  element of $\ccc$, then
 $ptp_n({\cal C}, c, \Theta) \neq ptp_n({\cal C}, a, \Theta)$ for each $n\geq 1$. This is because we allowed a query of the form
$y=c$, which  belongs to $ptp_1({\cal C}, c, \Theta)$, but not to  $ptp_1({\cal C}, a, \Theta)$.
\end{remark}

All the signatures under consideration are finite, so the number of possible conjunctive queries
with at most $n$ variables  is finite. In consequence the number of
  positive $n$-types is finite, for a given $n$.

Let us remark here that our positive $n$-types carry much less information  than the standard
first order types (in the sense of Geifman or Hanf).
Take for example a structure $\cal C$, over the signature $\Theta=\{R,E\}$, consisting of elements $a$,$b$,$c$,$d$,$e$,
and atoms $R(a,b)$, $R(a,c)$, $E(a,c)$, $E(d,e)$, $R(d,e)$. Then $ptp_2({\cal C}, a, \Theta)=$ $ptp_2({\cal C}, e, \Theta)$.
But the first order 2-types of $a$ and $e$ differ: consider for example the formula:
 $\psi(x)=\exists z,y \;  R(x,y)\wedge E(x,z) \wedge y\neq z$.
Then ${\cal C}\models \psi(a)$ but ${\cal C}\not\models \psi(d)$.

\subsection{How the finite structures are born}\label{born}

\begin{definition}\label{otypach}
Let  $d$ and $e$ be two elements  of $\cal C$. We define $d\equiv_n e$  if and only if
$ptp_n(\ccc, d, \Theta)=ptp_n(\ccc, e, \Theta)$.
\end{definition}

Notice that both the relation $\equiv_n$ and the structures $M_n(\ccc)$ (as defined below) depend on $\Theta$, and the signature should be
added as a parameter in Definitions \ref{otypach} and \ref{emen}. We will try to avoid confusion while keeping the notation light, but when really needed we will include the parameter, writing $M_n^\Theta(\ccc)$ instead of  $M_n(\ccc)$.

\begin{definition}\label{emen}
For a relational structure $\ccc$ define 
$M_n(\ccc)$ as a relational structure whose set of elements is $\ccc/ \equiv_n $, and such that $M_n(\ccc)\models R(<[a_i]_{\equiv_n}>_i)$ iff 
$\forall i~\exists b_i\in [a_i]_{\equiv_n}$ such that $\ccc\models R(<b_i>_i)$.

\end{definition}

In other words, the relations in $M_n(\ccc)$ are defined in the natural way, as  minimal (with respect to inclusion) relations such that the quotient mapping $q_n: \ccc\longrightarrow M_n(\ccc)$ is a homomorphism.

We usually imagine $q_n$ as a projection\footnote{''Projection'' in the geometric sense not the database sense.},
 so that the atoms in $M_n(\ccc)$ are  projections of atoms in $\ccc$.

Clearly each  $M_n(\ccc)$ is a finite structure.

\begin{lemma}\label{silnie}
If $q_n(d)=q_n(e)$ then $q_{n-1}(d)=q_{n-1}(e)$.
The structure $M_{n-1}(\ccc)$ is a homomorphic image of  $M_{n}(\ccc)$.
\end{lemma}

For the proof of the first claim notice that it follows from Definition \ref{ptp-def} that if the positive $n$-types of two elements are equal then their 
positive $(n-1)$-types are also equal. The second is an easy exercise in basic universal algebra. 

\piki{naduzycie}{\em The function $q_n$, as defined above, has $\ccc$ as its domain. It will however be 
convenient to be able to write $q_n(a)$ also for $a\in M_{n+1}(\ccc)$. In such a case
$q_n(a)$ will be defined  as $q_n(b)$, where $b\in \ccc$ is any element such that $q_{n+1}(b)=a$.
It follows from Lemma \ref{silnie} that the value of $q_n(a)$ does not depend on the choice of $b$.}

We defined a canonical way of building finite structures. 
But is there any chance that they really resemble the original infinite structure? 
What we are particularly interested in is  what happens to the positive $m$-types of elements of $\ccc$. 
Are they preserved by $q_n$? It is easy to see that we always have 
$ptp_m(\ccc, e,\Theta) \subseteq ptp_m(M_n(\ccc),q_n(e),\Theta)$. But can the inclusion be replaced with equality?
Is the positive $m$-type of $e\in \ccc $ always the same 
as the positive $m$-type of $q_n(e)$?\footnote{Notice that we use two natural numbers here: $n$,
which we imagine is big -- the bigger it is the more similar $M_n(\ccc)$ and $\ccc$ are,  and $m$ -- the smaller it is the  easier it is preserve the positive $m$-types.} 
Unfortunately this is not yet the case:

\begin{example}\label{oczywisty}

Let $\Sigma =\{ E\}$ and let  $\ccc$ be the set $\{a_0,a_1,a_2$ $\ldots\}$ with $E(a_i,a_{i+1})$ for each $i$.
Notice that the names of the elements $a_i$ are not part of $\Sigma$, so they are invisible for the inhabitants of the structure,
 and the positive $n$-types of  elements $a_i$ and $a_j$, with $i\neq j$, are equal if and only if $i,j\geq n$. 
(Actually, not only the positive types of $a_i$ and $a_j$ are equal, but even their $n$-Gaifman neighborhood are isomorphic.)
So $M_n^\Sigma(\ccc)$ is a structure with elements $\{b_0,b_1,b_2,\ldots b_n\}$, with  $E(b_i,b_{i+1})$ for each $i<n$
and with $E(b_n,b_n)$. Clearly, $q_n(a_n)=b_n$. But the positive 1-type of $b_n$ in $M_n(\ccc)$  contains the query $\exists y R(y,y)$, 
which is not in the positive 1-type of $a_n$ in $\ccc $.
\end{example}

\subsection{Colored structures}

We are not quite happy with the quotient structure we got in Example \ref{oczywisty}. Too many elements of $\ccc$ 
are identified, 
and  even a very small conjunctive query can easily see the difference between $\ccc$ and $M_n(\ccc)$.
But consider another example:

\begin{example}\label{oczywisty+kolory}
Let $\Sigma$ and $\ccc $ be like in Example \ref{oczywisty}. Let $\bar\Sigma =\Sigma \cup \{K_0, K_1,\ldots K_m\}$, where 
$K_0,K_1,\ldots K_m$ are unary predicates (colors) and let $\bar{\ccc}$ be like $\ccc $, but with each $a_i$ satisfying  also $K_{i \mod (m+1)}$.

Let $n>m$. 
Then the positive $n$-types of  elements $a_i$ and $a_j$, with $i\neq j$, are equal if and only if $i,j\geq n$ and 
$i=j \mod {m+1}$.
So $M_n^{\bar\Sigma}(\barccc)$ will be a structure with elements $\{b_0,b_1,b_2,\ldots b_{n+m}\}$, with  $E(b_i,b_{i+1})$ for each $i<n+m$ and with $E(b_{n+m},b_n)$. 
It is not hard to see that now 
$ptp_m({\barccc}, a, \Sigma)=$ $ptp_m( M_n^{\bar\Sigma}(\barccc) , q_n(a), \Sigma)$ for each element $a\in \barccc$. 
The positive $m$-types of the elements of $\ccc $
are preserved by the quotient operation. 

Notice however that the positive $(m+1)$-types 
are not preserved. This is because, unlike $\ccc$, the structure
 $M_n^{\bar\Sigma}(\barccc)$ contains a cycle of length $m+1$, which is easy to detect with a query with $m+1$ variables. 
If we want to preserve positive $m$-types for bigger numbers $m$ we need to use  more colors. 

Notice also that if we took $n<m$ then we would get 
 $ptp_m({\barccc}, a_{n}, \Sigma)\neq$ $ptp_m(M_n^{\bar\Sigma}(\barccc), q_n(a_{n}), \Sigma)$. This is because
$a_{n}$ would then be identified with all the elements $ a_{n+k{m+1}}$ for $k\in \mathbb{N}$, 
and therefore the query\vspace{1mm}\\
\hspace*{1mm}$\exists x_1\ldots x_{m-1} E(x_1,x_2) \wedge E(x_2,x_3)\wedge \ldots \wedge E(x_{m-1},q_n(a_{n}))$\vspace{1mm}\\
would be satisfied in $M_n^{\bar\Sigma}(\barccc)$, while the query \vspace{1mm}\\
\hspace*{5mm}$\exists x_1\ldots x_{m-1} E(x_1,x_2)\wedge E(x_2,x_3)\wedge \ldots \wedge E(x_{m-1},a_{n})$\vspace{1mm}\\
 is not satisfied in $\barccc$.
\end{example}

The last example motivates the following definitions:

\begin{definition}
 Each of the unary predicates $K_h^l$ for some
$h,l\in \mathbb N$  will be called a color, with the number $h$ being called the hue of the color and
the number $l$ being called its lightness.
 The set of all colors will be denoted as $\cal K$.
\end{definition}

So far we just defined an infinite set of unary predicates (with strange names, that we will need much later).
Now a definition of coloring. A  natural one:

\begin{definition}
For a structure $\ccc$ over a signature $\Sigma$ by a coloring of $\ccc$ we will mean a 
structure $\barccc$ over some finite signature $\bar\Sigma $ such that:

\begin{enumerate}\label{coloring}
\item $\Sigma \subset \bar\Sigma \subset \Sigma \cup \cal K$ 
\item $\barccc \upharpoonright  \Sigma = \ccc $
\item for each $a\in \barccc $ there is exactly one color $K\in \cal K$ such that $\barccc \models K(a)$.
\end{enumerate}

where $\barccc \upharpoonright  \Sigma$ is the structure  $\barccc $ restricted to the signature $\Sigma$.
\end{definition}

\subsection{Conservative structures}\label{types}

\begin{definition}\label{conservative-1}
Let  $\ccc$ be a structure 
and let $m,n\in \mathbb{N}$. We will say that a coloring  $\barccc$ of $\ccc $
is $n$-conservative up to size $m$ if:   

\piki{takimabyc}{  $\; ptp_m(\ccc, e,\Sigma) = ptp_m(M_n^{\bar\Sigma}(\barccc),q_n(e), \Sigma)$ 
for each $e\in \ccc$, where $\Sigma$ and $\bar\Sigma$ are like in  Definition \ref{coloring}.
}
\end{definition}

Being $n$-conservative up to size $m$ means that the positive 
$m$-types ({\bf with respect to the signature $\Sigma$}) 
of elements of $\barccc $ are preserved by the quotient mapping $q_n$ leading to the structure 
 $M_n^{\bar\Sigma}(\barccc)$.
So, for example, the coloring $\barccc$ from Example \ref{oczywisty+kolory} is $n$-conservative up to size $m$,
if only $n>m$, but is not  $n$-conservative up to size $m+1$ for any $n$. 

\begin{definition}\label{conservative-2}
 A structure $\ccc$  is ptp-conservative if 
 for each $m\in \mathbb{N}$ there exist $n\in \mathbb{N}$ and a coloring  $\barccc$ of $\ccc $, such that 
$\barccc$ 
is $n$-conservative up to size $m$.
\end{definition}

The following remark will be useful in Section \ref{odlozona}

\begin{remark}\label{jednostajne}
Consider a coloring  $\barccc$ of $\ccc $ and a number $m$. 
Suppose there is no such $n$ that $\barccc$ is  $n$-conservative up to size $m$. 
This means that for each $n\in \mathbb{N}$ there is a query $\exists \bar x\Psi_n(\bar x, y)$, with at most $m$ variables,
and an element $e\in \ccc$ such that $M_n(\barccc)\models\exists \bar x \Psi_n(\bar x, q_n(e))$ but 
$\ccc\not\models \exists \bar x \Psi_n(\bar x, e)$. 

But since there are only finitely many queries of at most $m$ variables, this implies that there is a query
$\Psi$ which is $\Psi_n$ for infinitely many numbers $n$.

Notice also that if $n'<n$ and $M_n(\barccc)\models\exists \bar x \Psi(\bar x, q_n(e))$ then 
$M_{n'}(\barccc)\models \exists \bar x \Psi(\bar x, q_{n'}(e))$. 

So, if there is no such $n$ that $\barccc$ is  $n$-conservative up to size $m$ then it must exist a 
single query $\Psi(\bar x,y)$, with $|x|<m$,
such that for every $n$ there is an element $e$ of $\ccc$ such that $M_n(\barccc)\models \exists \bar x\Psi(\bar x, q_n(e))$ but 
$\ccc\not\models \exists \bar x \Psi(\bar x, e)$.
\end{remark} 

\subsection{Further examples and remarks}

\begin{example} It is very easy to see that the structure $\ccc$ from Examples \ref{oczywisty} 
and \ref{oczywisty+kolory} is ptp-conservative. Given $m$
one just needs to define the coloring $\barccc$ using $m+1$ colors, like in Example \ref{oczywisty+kolory}, and take $n=m+2$.
Then $\barccc$ 
will be $n$-conservative up to size $m$.
\end{example}

\begin{example}\label{niekonserwatywne}
Let $\ccc$ be any infinite set with a total (irreflexive) order $E$. Then it is easy to see that $\ccc $ is not ptp-conservative. Actually, it  is impossible to find a  coloring $\barccc$ of $\ccc$ and a number $n$ such that 
 $\barccc$ is $n$-conservative up to size 1:
 whatever the coloring, there would be an element $e$ in $\ccc$ such that $M_n^{\bar\Sigma}(\barccc)\models E(q_n(e),q_n(e))$.
\end{example}

\begin{remark}\label{teoriazporzadkiem}
It is very important to see the role of the element $e$ in Definition \ref{conservative-1}. 
Condition \rpiki{takimabyc}, which says that each element of $\ccc$ keeps its positive type after 
the quotient operation, is {\bf strictly} stronger than:

\piki{nietaki}{for each conjunctive query $\Psi$ over $\Sigma$, 
with at most $m$ variables, $\ccc\models \Psi$ if and only if $M_n^{\bar\Sigma}(\barccc)\models \Psi$.
}

which says that no new positive $m$-types appear in $M_n^{\bar\Sigma}{\barccc}$. 

To see that, consider a theory $\cal T$ consisting of the rules:

$ E(x,y)\Rightarrow \exists z\; E(y,z) $

$E(x,y),E(y,z) \Rightarrow  E(x,z) $

and a database instance $D=\{E(a,a), E(b,c)\}$.

Let $\ccc=Chase(D, {\cal T})$. Then $\ccc$ satisfies condition \rpiki{nietaki}
(since, due to the presence of the atom $E(a,a)$, 
 all possible queries are  true in $\ccc$),
but is not ptp-conservative as it contains an infinite irreflexive total order (see Example \ref{niekonserwatywne}).
\end{remark}

Next remark explains what Definition \ref{conservative-2} is good for:

\begin{remark}\label{goodfor}
Imagine that we have some  theory $\cal T$ over a binary signature $\Sigma$, and
there is a existential TGD $\Psi$ in $\cal T$ of the form $\psi(\bar x,y)\Rightarrow \exists z \; R(y,z)$,  where $m$ is the number
of variables in $x$. Let $\ccc = Chase(D, {\cal T})$.
Clearly $\ccc \models \Psi$, so if for some $e\in \ccc$ it holds that $\exists \bar x \psi(\bar x, y)\in ptp_m(\ccc,e, \Sigma)$,
then there exists $d\in \ccc$ such that $\ccc\models R(e,d)$.

Suppose we now color  $\ccc $ and project it, using $q_n$,  creating  some finite structure $M_n(\barccc)$. 
 We would like 
to be sure that $M_n(\ccc)$ is still a model (or at least some sort of pre-model) of $\cal T$. 
So in particular we would like to be sure that 
$M_n(\ccc)\models \Psi$.  

But if  $\barccc$ was $n$-conservative up to  size $m$, then we can be sure 
that whenever we have an element $a=q_n(e)$ in $M_n(\barccc)$, such that $M_n(\barccc)\models  \exists \bar x\psi(\bar x,a)$ 
then also $\ccc\models \exists \bar x \psi(\bar x,e)$, which  implies that $\ccc\models \exists z\; R(e,z)$, which
implies  that  $M_n(\barccc)$ $ \models  \exists z\; R(a,z)$ (notice that what we use here is really Condition \rpiki{takimabyc} and that 
Condition \rpiki{nietaki} would not be strong enough)

So, if   $\barccc$ was $n$-conservative up to  size $m$, then $M_n(\barccc)$ is a model of $\Psi$, and if 
$\ccc$   is ptp-conservative then we can choose $m$ greater than the maximal size of the body of a existential TGD rule in 
$\cal T$  and be sure that there exists a coloring, and number $n$, leading to $M_n(\barccc)$  in which  all the existential TGDs
of $\cal T$ are satisfied. 
\end{remark}

We now know how to turn a  ptp-conservative Chase $\ccc$ of $\cal T$ into a finite structure $M_n(\barccc)$ satisfying all the existential TGDs in $\cal T$.  
But does it mean that $M_n(\barccc)\models {\cal T}$ then?  As the following example shows, not necessarily, even if $\cal T$ is BDD:

\begin{example}\label{jakniemozna} 
Consider the following BDD theory $\cal T$:

$E(x,y)\Rightarrow \exists z\; E(y,z)$

$E(x,y),E(x',y)\Rightarrow R(x,x')$

and a database instance $D=\{E(a,b)\}$.

Let $\ccc = Chase(D, {\cal T})$. Clearly, $\ccc$ is an infinite $E$-chain, with an atom $R(e,e)$ true for each $e\in\ccc$.
Whatever coloring we now use, the only $R$ atoms in $M_n(\barccc)$ will be the ones 
of the form $R(e,e)$. And whatever the coloring, there 
must be a triple of elements $a,b,c$ in $M_n(\barccc)$ such that $a\neq b$ and $M_n(\barccc)\models E(a,c),E(b,c)$,
which shows that $M_n(\barccc)$ is not a model of the plain datalog rule from $\cal T$.
\end{example}

Of course all the above definitions -- of types, of $\equiv_n$, of $M_n$ and of conservativity, 
make sense also when we consider any signatures, not just binary. But Remark \ref{goodfor} is not valid any more for such signatures,
which means that it is very hard to make sure that $M_n$ will  actually resemble a model of $\cal T$. 
We will be back to this point in Section \ref{why}.

\subsection{Very Treelike DAGs and the Main Lemma}

Most of the notions we defined so far apply to structures over any signature. 
But what we are really interested in in this paper are binary signatures.
They consist of some binary relations, some unary relations and constants.  
Structures over such signatures  can be in a natural way  seen as directed graphs with edges, and vertices, labeled with some finite 
number of labels (i.e. the names of the relations).
 Thanks to that we can use the language of graphs -- for example our 
infinite structures are usually (directed) trees or DAGs. 

%

We will concentrate  on Very Treelike DAGs:

\begin{definition}\label{poprzedniki}
For an element $e\in \ccc $ we define ${\cal P}(e)=\{e\}$  if $e\in \ccc\con $  and 
${\cal P}(e)=\{e\}\cup \{x\in \ccc\non : \ccc\models R(x,e) $ for some  $R\in \Sigma\; \}$ if $e\in \ccc\non $
\end{definition}

\begin{definition}\label{vtdag}
A structure $\ccc$ is called a Very Treelike DAG (VTDAG) if $\ccc\non $ is a DAG and:
\begin{itemize}
\item
 for each binary relation $R$ and each 
 $e\in \ccc\non$ there is at most one  $d\in \ccc\non$ such that $R(d,e)$;
\item
for each   $e\in \ccc\non$ if $d,d'\in {\cal P}(e)$ then $d\in {\cal P}(d')$ or $d'\in {\cal P}(d)$.

\end{itemize}
\end{definition}

The first condition says that each non-constant $e$ has at most one non-constant ''direct predecessor'' in each
binary relation. The second says that the set of  ''direct predecessors'' of $e$ is a (directed) clique. 

Each tree is trivially a VTDAG. In order to prove Theorem \ref{glownewintro}
it is  enough to restrict the attention to trees only. VTDAGs which are not trees will not be considered before 
Section \ref{dyskusja}.

%
%
%
%
%
%
%

The main tool in the proof of Theorem \ref{glownewintro} is:

\begin{lemma}\label{duzy}[The Main Lemma]\smallskip\\
Each VTDAG is ptp-conservative.
\end{lemma}

In Section \ref{dowod} we  use Lemma \ref{duzy} to prove Theorem \ref{glownewintro}.
Then, in Section \ref{odlozona}, we present a proof of Lemma \ref{duzy}.
Sections \ref{dowod} and \ref{odlozona} are independent and can be read in any order.


\section{From the Main Lemma  to  Theorem 2}\label{dowod}

\subsection{Hiding the query inside the theory}\label{hiding}

Nothing  complicated happens in this subsection. 
We are just making some simplifying (although without loss of generality) assumptions 
about the BDD theory under consideration. This will help us to keep the notations simpler in the rest of Section \ref{dowod}

For a binary BDD theory ${\cal T}_0$ and a conjunctive query $Q(\bar x, y)$ define 
a new theory ${\cal T}$ as ${\cal T}_0$ enriched with a new TGD:

\piki{newtgd}{
 $Q(\bar x, y) \Rightarrow \exists z \; F(y,z) $}

where $F$ is a new predicate symbol.
It is now easy to see  that, 
for any database instance $D$ such that $F$ does not occur in $D$,
a finite structure ${\cal M}$ such that
${\cal M}\models {\cal T}_0, D, \neg Q$ exists if and only if 
a finite structure ${\cal M}$ such that
 ${\cal M}\models {\cal T}, D, \neg F $ exists. This means that, in order to prove Theorem \ref{glownewintro}  it is enough to show:

\begin{theorem}\label{glowne}
For a binary BDD theory $\cal T$, containing a rule of the form  \rpiki{newtgd}, with predicate $F$ not occurring anywhere
else in $\cal T$, and for each database instance $D$, if $F$ does not occur in $Chase(D,{\cal T})$ then
there exists a finite structure $\cal M$ such that ${\cal M}\models D,{\cal T}$, without any atom 
of predicate $F$ occurring in $\cal M$.
\end{theorem}

From now on we assume that $\cal T$ is like in the assumptions of the above  Theorem. 
We also assume, in order to keep the notations simple, that:

\piki{warunkinat}{

\begin{itemize}
\item
the head  of each existential TGD in  $\cal T$ is of the form $\exists z\; R(y,z)$, which means that the 
witness, whose existence is demanded by the TGD, is the second argument of the predicate in the head;
\item 
if the predicate $R$ occurs as the head of some existential TGD in $\cal T$ then it does not occur as the head of any datalog rule in 
$\cal T$. We call such predicates TGPs -- tuple generating predicates.
\end{itemize}
}

We leave it for the readers as an exercise to see that  every $\cal T$ can be 
easily modified to satisfy \rpiki{warunkinat}, for the cost of some additional predicates and datalog rules, 
and this modification neither changes the BDD status of the theory nor its FC status. 

Hint: For each predicate $R$ in the signature introduce two new predicates $R'$ and $R''$.
Add to theory datalog rules $R'(x,y) \rightarrow R(x,y)$ and $R''(x,y) \rightarrow R(y,x)$.
Replace each head of an existential TGD which is of the form $\exists z\; R(y,z)$ by $\exists z\; R'(y,z)$
and each head of the form $\exists z\; R(z,y)$ by $\exists z\; R''(y,z)$.

\subsection{The structure ${\ccs}(D, {\cal T})$}

Let now $D$ be a database instance without atoms of $F$ and let $\Theta $ be the signature of $D$ and $\cal T$. 
Define $\Sigma \supseteq \Theta $ as a new signature  which contains, 
apart from the relations and constants from $\Theta$, a name for each element in $D$. Why do we prefer the elements
of $D$ to be named? Because we want to be sure that their positive types in $Chase(D, {\cal T})$ 
differ, and, in consequence, that they
 remain distinct after a quotient operation (see Remark \ref{pozna}).

Now we are going to define the structure to which the techniques of Section \ref{teoria} will be applied. Since we want to make use of Lemma \ref{duzy}, this structure must be a tree (or at least a VTDAG). 
And of course we cannot expect $Chase(D, {\cal T})$ to be a VTDAG.
 
\begin{definition}\label{kora}
By $\ccs(D,{\cal T})$ (or just $\ccs$,  as the context is always clear) we mean 
the substructure of $Chase(D, {\cal T})$ consisting of all the elements of $Chase(D, {\cal T})$, all 
 the atoms in $D$ and all the atoms of the TGPs. We understand that $\ccs$ is a structure over the signature $\Sigma$. 
\end{definition}

We will call the atoms in $\ccs$ {\em skeleton atoms}, as we imagine $\ccs$ as a sort of a skeleton of $Chase(D, {\cal T})$.
 The atoms of $Chase(D, {\cal T})$ which are not in $\ccs$ will be called {\em flesh atoms}. So the flesh atoms are the ones created in the process of chase by the datalog rules.

It follows easily from  \rpiki{warunkinat} that:

\begin{lemma}
\begin{enumerate}
\item[(i)] The graph $\ccs\non $ is acyclic;

\item[(ii)] the in-degree of any element of $\ccs\non$  is 1;

\item[(iii)] $\ccs\non$ is a forest;

\item[(iv)] the degree of the elements of $\ccs\non$ is bounded by $|\Sigma|+1$;
\end{enumerate}
\end{lemma}

Remember that  all the elements of $D$ are constants from $\Sigma$, so they are not in $\ccc\non$. 

\proof For the proof of (i) and (ii) notice that the only way a TGP atom $R(a,b)$ 
can be created is to be created together with a new element $b$. Acyclicity follows from the fact that
$b$ is always a "younger" element of $Chase(D, {\cal T})$ than $a$. The claim (iii) follows from (i) and (ii). 
Finally, (iv) follows from the fact, that the chase we consider is a non-oblivious one,
so for any fixed $a\in \ccs$ and for any TGP $R$ from $\Sigma$ at most one $b\in  \ccs$ can exist such that 
$\ccs\models R(a,b)$. \eop


Let us now think of $\ccs$ as of a  new database instance:

\begin{lemma}\label{nicponadto}
$ Chase(D, {\cal T}) \models Chase(\ccs, {\cal T})$. 
In particular, $\dom (Chase(\ccs, {\cal T}))=\dom (Chase(D, {\cal T}))=\dom (\ccs)$
\end{lemma}

\noindent
{\bf Proof:}

 It is an easy lemma. It is enough to show, by induction, that for each natural $n$:

\piki{jednagw}{
$Chase(D, {\cal T})\models Chase^n(\ccs, {\cal T}) $.}

Clearly, $  Chase(D, {\cal T}) \models Chase^0(\ccs, {\cal T}) $, by our definition of $\ccs$.
Suppose that \rpiki{jednagw} is true for some $n\in \mathbb{N}$, and 

\piki{niewprost7}{let $R(b,b')$ be an atom which is not true in $Chase^{n}(\ccs, {\cal T})$}
but true in $Chase^{n+1}(\ccs, {\cal T})$. In order to prove that this implies 
$Chase(D, {\cal T})\models R(b,b')$ notice that one of the following two possibilities must 
hold:

\piki{druga}{there is a datalog rule of the form   $\Psi(\bar x, y, z)\Rightarrow   R(y,z)$
in $\cal T$ and  $ Chase^n(\ccs, {\cal T}) \models \Psi(\bar a, b, b') $ for some elements $\bar a, b, b'$, or} 

\piki{pierwsza}{there is TGD  of the form  $\Psi(\bar x, y)\Rightarrow \exists z \; R(y,z)$
in $\cal T$ and  $ Chase^n(\ccs, {\cal T}) \models \Psi(\bar a, b) $ for some elements $\bar a,b$.} 

In both cases it follows from the inductive assumption  that $Chase(D, {\cal T})\models \Psi(\bar a, b, b') $ 
(resp. $\Psi(\bar a, b) $).
Now, if  \rpiki{druga} then $Chase(D, {\cal T})\models R(b,b')$ follows simply from the fact that all the 
rules of $\cal T$ are satisfied in $Chase(D, {\cal T})$. Similarly, if \rpiki{pierwsza} then there exists $b'$
such that  $Chase(D, {\cal T})\models R(b,b')$.  
By the definition of $\ccs$, we  have that
$\ccs\models R(b,b') $, so also $ Chase^n(\ccs, {\cal T})\models R(b,b')$, which contradicts \rpiki{niewprost7}   \eop\bigskip

It is very easy to see that 
also $ Chase(\ccs, {\cal T})\models Chase(D, {\cal T})$, so we get that $ Chase(D, {\cal T}) = Chase(\ccs, {\cal T})$.

The idea behind the Lemma is that while $\ccs$ is a simple structure -- simple enough to be ptp-conservative -- still 
not only it contains all elements of $Chase(D, {\cal T})$ but also the complete information about the relations
between elements of $Chase(D, {\cal T})$ that need a witness and the needed witnesses. Thanks to that $Chase(D, {\cal T})$
can be rebuilt, starting from the skeleton $\ccs$, in a process of a (non-oblivious) chase that only triggers datalog
rules, but never the existential TGDs. Notice that this would no longer be true if a single atom $R(a,a')$ was removed from $\ccs$
(even if the elements $a$,$a'$ were kept, as arguments of some other atoms). This is because at some point a TGD with the head
$ \exists x\; R(a,x)$ would be triggered, and a {\bf new} element $a''$ would be created.



\subsection{Proof of Theorem \ref{glowne}}
Recall that for a BDD theory $\cal T$ and query $\Psi$ by $\Psi'$ we mean
the positive first order rewriting of $\Psi$, which means that $\Psi'$ is such a query (a union of conjunctive queries), that
for each database instance $D$ it holds that $Chase(D,{\cal T})\models \Psi \;\Leftrightarrow\; D\models \Psi'$. 

Let $\kappa =\max\{|Var(\Psi')|: \Psi\Rightarrow \psi$ is a rule in $ {\cal T}\}$. In other words,
$\kappa $ is the maximal number of variables in a query being a positive first order rewriting of a body
of some rule of the theory $\cal T$. By Lemma \ref{duzy} there exists a coloring $\barccs$ of $\ccs$
and $\eta\in \mathbb{N}$ such that $\barccs$ is $\eta$-ptp conservative up to the size $\kappa $, which means that 
the elements of $M^{\bar\Sigma}_\eta(\barccs)$ have the same positive $\kappa $-types over $\Sigma$ as their counter-images in $\ccs$.

Now there are five structures one should imagine:

\begin{itemize}
\item[(i)] $\ccs $  

\item[(ii)] Chase$(D,{\cal T})$ = $Chase(\ccs,{\cal T}) $ 

\item[(iii)] $M_\eta^{\bar\Sigma}(\barccs)$ 

\item[(iv)] Chase$(M_\eta^{\bar\Sigma}(\barccs),{\cal T})$ 

\item[(v)] $q_\eta($Chase$(D,{\cal T}))$ 

\end{itemize}

The first two of them were already introduced in this Section. The third is the result of the quotient operation
applied to $\barccs$ -- something we discussed in Section \ref{teoria}. Since 
$\barccs$ is $\eta$-ptp conservative up to the size $\kappa $, we know that $M_\eta^{\bar\Sigma}(\barccs)$ is a model
for all existential TGDs in $\cal T$ (see Remark \ref{goodfor}). But, as we saw in Example \ref{jakniemozna}, we cannot be sure 
that $M_\eta^{\bar\Sigma}(\barccs)\models {\cal T}$, as some datalog rules from $\cal T$ may be false in 
$M_\eta^{\bar\Sigma}(\barccs)$. So to get a model of $\cal T$ we apply chase to $M_\eta^{\bar\Sigma}(\barccs)$,
which leads to our fourth structure, Chase$(M_\eta^{\bar\Sigma}(\barccs),{\cal T})$. So far we know nothing about 
this structure, in particular we do not even know whether Chase$(M_\eta^{\bar\Sigma}(\barccs),{\cal T})$ is finite.

The fifth structure, $q_\eta($Chase$(D,{\cal T}))$ is only needed in example \ref{zbedny} which we hope explains some
issues concerning Chase$(M_\eta^{\bar\Sigma}(\barccs),{\cal T})$. 
{\bf If} you  feel you not need more explanations {\bf go directly to Lemma \ref{wystarczy}}. 
 $q_\eta($Chase$(D,{\cal T}))$ is  defined as:


\begin{itemize}
\item $\dom(q_\eta($Chase$(D,{\cal T}))) = \dom (M_\eta^{\bar\Sigma}(\barccs))$;

\item relations are defined in $q_\eta($Chase$(D,{\cal T}))$ as the minimal relations such that 
$q_\eta$, understood as a mapping from Chase$(D,{\cal T})$ to $q_\eta($Chase$(D,{\cal T}))$, is a homomorphism.
\end{itemize} 

So while the relations  $M_\eta^{\bar\Sigma}(\barccs)$  are defined as projections of the 
skeleton relations, the relations  $q_\eta($Chase$(D,{\cal T}))$  are projections of both, the skeleton and the flesh atoms. 

One can see that
Chase$(M_\eta^{\bar\Sigma}(\barccs),{\cal T}) \models q_\eta($Chase$(D,{\cal T}))$.
Indeed, any atom in $q_\eta($Chase$(D,{\cal T}))$ which is not in $M_\eta^{\bar\Sigma}(\barccs)$ 
is a projection of some flesh atom in 
Chase$(D,{\cal T})$. This last atom must have been proved by some derivation in Chase$(D,{\cal T})$. But a projection of
a  valid derivation from Chase$(D,{\cal T})$ is a valid derivation in 
 Chase$(M_\eta^{\bar\Sigma}(\barccs),{\cal T})$. 

At this point it would be reasonable to conjecture that maybe
$q_\eta($Chase$(D,{\cal T})) =$  Chase$(M_\eta^{\bar\Sigma}(\barccs),{\cal T})$.
But this is not always the case, as the following example shows:

\begin{example}\label{zbedny}
Let $\cal T$ and $D$ be like in Example \ref{jakniemozna}. Let $\ccc = Chase({\cal T},D)$.  
  Now $\ccs$ is the structure $\ccc$
from    Example \ref{oczywisty+kolory}. Let $m$ be the number of colors,  $n>m$ and let $\barccs $ be a coloring of
 $\ccs$, like in Example \ref{oczywisty+kolory}. 
Now, the only $R$ atoms in $q_n(M_n^{\bar\Sigma}(\barccs))$ are atoms of the form $R(a,a)$ for some $a\in M_n^{\bar\Sigma}$.
But it is easy to see that Chase$(M_n^{\bar\Sigma},T)\models R(b_{n-1}, b_{n+m})$ (where $b_{n-1}$, $b_{n+m}$ are again
 like in Example \ref{oczywisty+kolory}). 
\end{example}

In the last example  an atom was derived in  Chase$(M_\eta^{\bar\Sigma}(\barccs),$ ${\cal T})$, which 
was not a projection of any flesh atom. The meaning of the next 
lemma is that while,
in the process on chase on $M_\eta^{\bar\Sigma}(\barccs)$, 
some  datalog derivations can arise not being  projections of datalog derivations 
in chase on $\ccs$, still (like in Lemma \ref{nicponadto}) no existential
TGDs will  be used, and no new elements will be created.
To be more precise:\vspace{1mm}

\begin{lemma}\label{wystarczy} \hspace{-0.1mm}$\dom($Chase$(M_\eta(\barccs),{\cal T}))= \dom(M_\eta(\barccs))$
\end{lemma}

This proof is not very long but we believe it is quite tricky. 
It is here where things really happen: Lemma \ref{duzy} meets the assumption that $\cal T$ is BDD. 

\noindent
{\bf Proof of Lemma \ref{wystarczy}:}

 Suppose 
$\dom(Chase(M_\eta(\barccs),{\cal T}))\not\subseteq \dom(M_\eta(\barccs))$

and let $j$ be the smallest natural number such that

\begin{center}
$\dom(Chase^{j+1}(M_\eta(\barccs),{\cal T}))\not\subseteq  \dom(M_\eta(\barccs))$.
\end{center}

This means that there is a rule $\Psi(\bar x, y)\Rightarrow \exists z\; R(y,z)$ in $\cal T$
and elements $\bar a, b$ of $Chase^{j}(M_\eta(\barccs),{\cal T})$ such that 

$Chase^{j}(M_\eta(\barccs),{\cal T})\models \Psi(\bar a, b)$, but 

\piki{ostatni}{
 $Chase^{j}(M_\eta(\barccs),{\cal T})\not\models \exists z\; R(b,z)$.}

But if $Chase^{j}(M_\eta(\barccs),{\cal T})\models \Psi(\bar a, b)$ then of course also\\
$Chase(M_\eta(\barccs),{\cal T})\models \Psi(\bar a, b)$. Since all of $\bar a $ and $b$ are elements of $M_\eta(\barccs)$
we get (using the fact that $\Psi'$ is the positive first order rewriting of $\Psi$)
that $M_\eta(\barccs)\models \Psi'(\bar a, b)$ or, in other words, $M_\eta(\barccs)\models  \Psi'(\bar x, b)$. 
Let $c$ be any such element of $\ccs$ that $q_\kappa (c)=b$. Since the size of $\Psi'$ is at most $\kappa $, 
by Lemma \ref{duzy} we get  that
$\ccs \models  \Psi'(\bar x, c)$. Using again the fact that $\Psi'$ is the positive first order rewriting of $\Psi$
we get that $Chase(\ccs,{\cal T})\models   \Psi(\bar x, c)$. 

$Chase(\ccs,{\cal T})$ is of course a model of $\cal T$, so there must be an element $d$ such that
$ Chase(\ccs,{\cal T})\models R(c,d)$. It follows from Lemma  \ref{nicponadto} that $d\in \ccs$ and, in consequence,
$e=q_\eta(d)\in M_\eta(\barccs)$. But this implies that $M_\eta(\barccs)\models R(b,d)$ 
which contradicts \rpiki{ostatni}.\eop

As we show in Section \ref{why},
there is no  hope to have anything similar to Lemma \ref{wystarczy}  in the general (non-binary) case.

We are ready to present the
{\bf proof of  Theorem \ref{glowne}:}

In order to prove Theorem \ref{glowne} (and, in consequence, Theorem \ref{glownewintro})
we need  to show a finite model of $D$ and  $\cal T$ without any atom
of the predicate symbol $F$. The structure $Chase(M_\eta(\barccs),{\cal T})$ is clearly a model of $D, \cal T$. It 
follows from the Lemma \ref{wystarczy}  that its domain is exactly the domain of $M_\eta(\barccs)$, so it is finite. 

Since no atom of the relation $F$ occurs in $\ccs$ there is also no such atom  in $M_\eta(\barccs)$. So the only way any such atom could appear in $Chase(M_\eta(\barccs),{\cal T})$
would be to derive it in the process of chase. But the only rule that derives $F$ is a existential TGD which demands a new element,
and no such rule could have been used, due to Lemma \ref{wystarczy}.\eop



\section{Proof of the Main Lemma}\label{odlozona}


Fix a VTDAG $\ccc$ and a natural number $m$.
 Let $\Sigma$ be the signature of $\ccc$.
To prove Lemma \ref{duzy} we need to find $n\in \mathbb{N}$ and a coloring $\barccc$ of $\ccc$   
such that $\barccc$ is $n$-conservative up to the size $m$.

First let us define the coloring:

\begin{definition}
For $e\in\ccc$ let ${\cal P}(e)$ be like in Definition \ref{poprzedniki}.

\begin{itemize}

\item
For  $e\in \ccc\non$ let  ${\cal P}_0(e)= {\cal P}(e) $.

\item 
For $e\in \ccc\non$ let ${\cal P}_k(e)=\bigcup_{a\in {\cal P}_{k-1}(e)} P(a)$

\end{itemize}
\end{definition} 

\begin{definition}\label{kolorowanie}
A coloring  $\barccc$ of  $\ccc$ will be called natural if it satisfies  the following conditions:

\begin{itemize}
\item if $e,e'\in \ccc$ are such that $e'\in {\cal P}_m(e)$ and if\\ $\barccc\models K^l_h(e), K^{l'}_{h'}(e')$ 
then $h\neq h'$;

\item if $e,e'\in \ccc$ are such that   $\barccc\models K^l_h(e), K^{l}_{h'}(e')$ then\\ 
$\ccc \restrictedto ({\cal P}(e)\cup \ccc\con) $ and $\ccc \restrictedto ({\cal P}(e')\cup \ccc\con) $ are isomorphic. 
\end{itemize}

\end{definition}

 It is easy to see that for each VTDAG $\ccc$ there exists a natural coloring $\barccc$. From now on 
by 
$\barccc$ we will mean a fixed natural coloring of $\ccc$. 


By Remark \ref{jednostajne} 
the proof of Lemma  \ref{duzy}  will be finished when we show:

\begin{lemma}\label{klaim}
For each query $\Phi(\bar x,y)$ over $\Sigma$, with $|\bar x|<m$,  there exists 
$n\in \mathbb{N}$ such that for each element $e\in \ccc$:
 
$M_n(\barccc) \models \exists \bar x \Phi(\bar x,q_n(e)) $ if and only if  $\ccc\models\exists \bar x \Phi(\bar x,e) $
\end{lemma}

{\bf
\underline{~~~~~~~~~~~~~~~~~~~~~~~~~~~~~~~~~~~~~~~~~~~~~~~~~~~~~~~~~~~~~~~~} 

Proof of Lemma \ref{klaim} begins here.~~~~~~~~~~~~~~~~~~~~~~~~~~~~~~~~~ 

\underline{It will take till the end of Section \ref{odlozona} to finish.~~~~~}
}

First of all notice that if the Lemma \ref{klaim} was false, then there would exists a {\bf counterexample}  --
a conjunctive  query $\Phi(\bar x, y)$  such that:

($\clubsuit $) for each $n\in\mathbb{N}$ 
there exists  an element $e_n^\Phi$ of $\ccc$ and a valuation $\sigma_n^\Phi:Var(\Phi)\rightarrow M_n(\barccc)$, with 
$\sigma_n^\Phi(y)= q_n(e_n^\Phi)$, such that $M_n(\barccc)\models \sigma_n^\Phi(\Phi)$
and $\ccc\not\models \exists \bar x \Phi(\bar x, e_n^\Phi)$.

Each time we will say that query $\Phi$  is a counterexample we will think that is satisfies  condition ($\clubsuit $). 


By a colors statement we will mean a query of the form:

\begin{center}
 $ \bigwedge_{z\in Var(\Phi)} K_{h_z}^{l_z} $ 
\end{center}

 where 
$K_{h_z}^{l_z}$ is any of the possible colors from $\cal K$. 
By a color closure of $ \Phi $ we will mean any query of the form
$ \Phi \wedge \Upsilon$, where $\Upsilon$ is a colors statement. Of course there are finitely many colors statements, 
and so there are finitely many possible color closures of $\Phi$. A query which is a color closure of some other query 
 will be called color closed. 

\begin{lemma}\label{zkolorami}
\begin{itemize}
\item[(i)]
 Let $\Phi$ be a counterexample.
Then for each $n$ there is a query $\Phi_c$, being a color closure of $\Phi$, such that 
$M_n(\barccc)\models \sigma_n^{\Phi}(\Phi_c)$
and $\barccc\not\models\exists \bar x \Phi_c(\bar x, e_n^{\Phi})$. 


\item[(ii)]  For each counterexample $\Phi$  there exists  a color closure  $\bar\Phi$ of $\Phi$, which also is a counterexample.

\item[(iii)] If there exists a query $\Phi$ being a color closed counter-example, then there also exists another 
color closed counterexample $\Psi$ such that for each constant $c$ from $\Sigma$, for each variable $z\in Var(\Psi)$ 
and  for each $n\in \mathbb{N}$ there is $\sigma^\Psi_n(z)\neq c$, where $\sigma_n^\Psi$ is as  $(\clubsuit)$. We will say that
counterexample $\Psi$ avoids constants.
\end{itemize}
\end{lemma}

\proof ~(i) The elements $\sigma_n^{\Phi}(z)$, where $z\in Var(\Phi)$ have some colors. Adding to $\Phi$ a statement asserting that 
they have the colors they really have will not make the new query $\sigma_n(\Phi_C)$ less true in $M_n(\barccc)$ 
than $\sigma_n^{\Phi}(\Phi)$ was. 

On the other hand,  $\exists \bar x \Phi(\bar x, e_n)$ was false in $\ccc$ already before the color statement was added and adding more constraints never
makes a query more true.

(ii) Use (i) and an argument like in  Remark \ref{jednostajne}.
 
(iii) Suppose  $\Phi$  is a color closed counterexample and $\sigma^\Phi_n(z)$ $=c$ for some
constant $c\in\Sigma$, some variable $z\in Var(\Psi)$ 
and  some $n\in \mathbb{N}$. 
By the definition of natural coloring, the color of $c$ is unique in $\ccc\con$ and thus
the equality $\sigma^\Phi_n(z)= c$ must hold for each $n$, and thus 
$\Psi$ being the result of replacing
each occurrence of $z$ in $\Phi$ by $c$ is also a  counterexample.\eop

We are now going to view queries as graphs. What we mean here is  
 a sort of Gaifman  graphs, where vertices 
are the variables in the query and the edges are the atoms of the query. As we only have binary and unary atoms, we can
in a natural way see each query as a directed (labeled) graph. 
Concerning the constants in the query, they are not understood to be vertices in the graph,
and it is  good to think that an atom of the form $R(a,x)$ in a query, where $a$ is a constant and $x$ is a variable, 
is just a unary predicate, telling us something about $x$ alone. Notice that atoms of the form $R(a,b)$ in a query, where both 
$a$ and $b$ are constants, are irrelevant from the point of view of Lemma \ref{klaim}, as
the part of $\ccc$ consisting of the constants remains unchanged after our projections. 

Now our plan of the proof of Lemma  \ref{klaim} is as follows.
We want to show that no query is an avoiding constants color closed counterexample. 
So first we will notice (Lemma \ref{acykliczne} and Lemma \ref{skier-cykl}) 
that neither a query  being an undirected tree, nor a query containing a 
directed cycle can ever be 
a counterexample. At this point we will know that if there is any avoiding constants color closed counterexample $\Phi$
then $\Phi $ must contain an undirected cycle (but not a directed one). But then, in Lemma \ref{niesk-cykl} 
we show that if such a $\Phi$ existed, then also another counterexample would exist, 
being a tree or  containing a directed cycle. That would however
 contradict Lemma \ref{acykliczne} and Lemma \ref{skier-cykl}. 

Proofs of Lemma \ref{acykliczne} and Lemma \ref{skier-cykl} are  
easy. Proof of Lemma \ref{niesk-cykl}, where we deal with queries 
containing an undirected cycle, is much more complicated. A technique of normalization of queries is used there,
 which we find
to be the deepest idea of this paper (we also 
employ this technique, in different context, in  \cite{GM13}, where it is called {\em second little trick}). Why are 
the undirected cycles in the query so much harder to deal with than  directed ones? The answer is 
in:


\begin{example}\label{cykle}
Let a theory $\cal T$ consist of the  rules:

$F(x,y)\Rightarrow \exists z\; F(y,z)$\hfill $F(x,y)\Rightarrow \exists z\; G(y,z)$

$G(x,y)\Rightarrow \exists z\; F(y,z)$\hfill $G(x,y)\Rightarrow \exists z\; G(y,z)$

 Let $D=\{F(a,b)\}$ and let $\ccc$ be $Chase(D, {\cal T})$, which means that
$\ccc $ is an infinite tree, where each element has exactly two successors. Or, in other words,
$\ccc$ consists, except from $a$ and $b$, of all the elements $w(b)$, where
$w\in\{f,g\}^*$. Let $\barccc$ be a natural coloring of $\ccc$.

Let  $e_1,e_2$ be two elements of $\ccc$ of the form
$e_1=vfw(b)$,  $e_2=vgw(b)$, where $v,w\in\{f,g\}^*$ and where 
$|v|=n-1$.  

 Then $a_1=q_n(e_1)\neq q_n(e_2)=a_2$ are two distinct elements of
$M_n(\barccc)$ -- the length of $v$ is not big enough to hide the slight difference in the positive types of $e_1$ and $e_2$.
 Of course also $a_3=q_n(f(e_1))\neq q_n(g(e_1))=a_4$ are two distinct elements 
(each of them distinct than $a_1$ and $a_2$). But $a_3=q_n (f(e_2))$ and $a_4 =  q_n(g(e_2))$.
This means that the atoms $F(a_1,a_3), F(a_2,a_3), G(a_2,a_4), G(a_1,a_4)$ are all true in $M_n(\barccc)$, and 
so there is an undirected cycle in $M_n(\barccc)$ consisting of 4 distinct elements.
\end{example}


As we saw in Example \ref{oczywisty+kolory}, by  using  coloring we can easily make sure that there are no small
directed cycles in $M_n({\barccc})$. But we cannot rule out small undirected new 
(not present in $\barccc$) cycles in $M_n({\barccc})$. So we need to prove that, while the new cycles exist,
 no small query can actually notice them.

\begin{lemma}\label{acykliczne}
Let $n\geq m$. Then for each element $e\in \ccc$ and each query 
$\Psi(\bar x, y)$, which  is an undirected tree:\\ 
$\Psi \in ptp_m({\barccc}, e, \bar\Sigma)$ if and only if  $\Psi \in ptp_m(M_n({\barccc}),q_n(e),\bar\Sigma)$.

\end{lemma}

It of course follows from the lemma that no query being an undirected tree can be a counterexample.

\begin{lemma}\label{skier-cykl}
Let $n\geq m$. Suppose $\Phi$ is a  query containing a directed cycle, by which we mean  a  sub-query of the form:\\
 $R_1(x_1,x_2),R_2(x_2,x_3),\ldots R_{k-1}(x_{k-1},x_k), R_{k}(x_{k},x_1)$\\
 where $k<m$ and $R_i$ are relation symbols from $\Sigma$.
 Then  $M_n({\barccc})\not\models \Phi$.
\end{lemma}

Clearly, as being a counterexample means, among other conditions, being true in $M_n(\barccc)$,
 the lemma  implies that  $\Phi$, containing a directed cycle, never is a counterexample.

{\bf Proof of Lemma  \ref{acykliczne}:}\vspace{-1mm}

Call a query  bad, if it is a counterexample\footnote{This is because the word ''counterexample'' is already being 
used in another context, 
and we want to avoid confusion}  to Lemma  \ref{acykliczne}.

Suppose there exist bad queries and  consider the smallest 
(with respect to the number of variables) of them. Call this query $\Psi$.

First notice that $y$ occurs only in one binary atom  in $\Psi$. Otherwise (since $\Psi$ is a tree) $\Psi$ could be
seen as a conjunction of smaller queries sharing only the variable $y$, and one of those queries would need to be 
a smaller bad query.

This means that the query 
$\Psi$ is either $(i)$ of the form $Q(x,y)$, where
 $Q(x,y)=R(x,y)\wedge \tau(x)\wedge\tau'(y)$ or $Q(x,y)=R(y,x)\wedge \tau(x)\wedge\tau'(y)$ for some relation symbol $R$
and unary queries $\tau$,$\tau'$,
or $(ii)$ of the form $Q(y,x_1)\wedge \Psi_0(x_1, \bar z)$, where $Q$ is as above 
 and $\bar z$ contains all the variables in $\bar x$ except from $x_1$. Notice that $\Psi_0$ is smaller
than $\Psi$.

The case $(i)$ is of course very easy. Remember that only elements of $\barccc$ 
the same unary type can be identified by our quotient operations.
So if $M_n(\barccc)\models  Q(x,q_n(e))$ then there must exist $d',e'\in \ccc$,
such that $\ccc \models Q(d',e')$ and $e\equiv_n e'$. But that implies, by definition of the relation $\equiv_n$, that
 there exists $d\in \ccc$ such that 
$\ccc \models Q(d,e)$. 

For the case (ii) suppose that $M_n(\barccc)\models  Q(q_n(e),x_1)\wedge \Psi_0(x_1, \bar z)$, and 
let $\sigma: \bar x \rightarrow  M_n$ be the satisfying valuation. 

$M_n(\barccc)\models Q(q_n(e),\sigma_n(x_1))$ means that there exist elements $e',d'\in \ccc$ such that 
$e\equiv_n e'$, $q_n(d')=\sigma(x_1)$ and $\ccc\models Q(e',d')$. 

Notice that
$M_n(\barccc) \models   \Psi_0(q_n(d'),\bar z)$. Since $\Psi_0$ is smaller than $\Psi$ it cannot be bad.
So we get that $\barccc \models   \Psi_0(d',\bar z)$. But that implies that $\barccc \models  \Psi(e',\bar x)$.
Now, since $e\equiv_n e'$ and  $m\leq n$, we get 
$\barccc \models \Psi(e,\bar x)$, but this contradicts the assumption that query $\Psi$ was bad.\eop

{\bf Proof of Lemma  \ref{skier-cykl}:}\vspace{-1mm}

 Suppose $\Phi$ is like in the Lemma and $M_n({\barccc})\models \Phi$. This means that there 
exist elements $b_1, a_2$, $b_2, a_3\ldots b_k,a_1$  of $\barccc$ such that $a_i\equiv_n b_i$ and $\barccc\models R_i(b_i,a_{i+1})$ 
for each $i$. Let $c_1=b_1$. Suppose $c_i$ is already defined, and $c_i\equiv_{n-i+1} b_i$. There exists a non-constant 
element of $\barccc$, namely $a_{i+1}$ such that $\barccc\models R_i(b_i,a_{i+1})$. Since  $c_i\equiv_{n-i+1} b_i$ and $n-i+1>0$, 
there exists (a unique) element $c_{i+1}\in \barccc$ such that $R_i(c_i,c_{i+1})$. By Lemma \ref{wygodny}
we get that $c_{i+1}\equiv_{n-i} b_{i+1}$.

But that means that $c_k\equiv_1 b_k$, which implies that there must be an element $d\in \barccc$ such that 
$\barccc\models R_k(c_k,d)$, where the color of $d$ is the same as the color of $c_1$, so
$c_1,c_2,\ldots d$ form a directed path in $\barccc$, of length not greater than $m$, joining two elements of the same color. This 
contradicts the definition of natural coloring. \eop

Now, Lemma \ref{klaim} follows from Lemma \ref{zkolorami}, Lemma \ref{acykliczne}, Lemma \ref{skier-cykl} and from the following:

\begin{lemma}\label{niesk-cykl}
If there exists a color closed counterexample $\Phi$ which avoids constants and which
 contains an undirected cycle  then
there exists also a counterexample being a tree or a counterexample containing a directed cycle.
\end{lemma}


\subsection{Proof of Lemma \ref{niesk-cykl}}\label{koniecdowodu}

Consider a query $\Psi(\bar x,y)$ which contains an undirected cycle, which is not a directed cycle. Then $\Psi$
 must be of the form:

($\heartsuit$) $R_1(z',z) \wedge R_2(z'',z)\wedge \psi(\bar x,y)$

for some relations $R_1,R_2\in \Sigma$ and some $z,z',z''$\hspace{-1mm}$ \in Var(\Psi)$.

\begin{lemma}[Normalization of queries]\label{normali}~~\\
If any color closed, avoiding constants, query $\Psi(\bar x,y)$ of the form $(\heartsuit)$   
is a counterexample, then there is a binary relation $P\in\Sigma$ such that
one of the following queries is also a color closed, avoiding constants, counterexample:

\begin{itemize}
\item $\psi(\bar x,y)\wedge R_1(z',z)\wedge z'=z'' $
\item $\psi(\bar x,y)\wedge R_1(z',z)\wedge P(z'',z')$
\item $\psi(\bar x,y)\wedge R_2(z'',z)\wedge P(z',z'')$
\end{itemize}

\end{lemma}

To see how Lemma \ref{normali} implies Lemma \ref{niesk-cykl},  {\bf while}
$\Psi$ is a counterexample of the form ($\heartsuit$)  {\bf do} replace it with another counterexample, the one whose existence is
assured by Lemma \ref{normali}. The only way to leave the while-loop is to produce a counterexample which 
is a tree or  contains a directed cycle. So it is enough to prove that the while-loop indeed terminates.

 If  the 
first  possibility from the Lemma is used as the replacement, 
then the new query has less variables than the old one (since adding
an equivalence of variables is the same as unifying the variables). But the last two possibilities do not decrease the number of variables.
So aren't they going to be applied forever? 
Consider the following measure of the size of a query:

\begin{center}
Measure($\Phi$)= $\Sigma_{x\in Var(\Psi)}\; occ(x)smaller(x)$
\end{center}

where $occ(x)$ is the number of the occurrences of variable $x$ in $\Psi$ and $smaller(x)$ is the number of 
variables from which $x$ is reachable by a directed path in the graph of the query. It is easy to see that Measure($\Psi)$
is a natural number which decreases each time Lemma \ref{normali} is applied. 

Before we prove Lemma \ref{normali} notice that the first condition in Definition \ref{vtdag} implies:

\begin{lemma}\label{wygodny}
Suppose $a,b,c,d$ are non-constant elements of $\barccc$, such that $\barccc \models R(a,b),R(c,d)$
for some relation $R\in\Sigma$. Then $b\equiv_n d$ implies $a\equiv_{n-1} c$.   
\end{lemma}

\proof ~Suppose there was a query $\psi(\bar x, y)$, with $|\bar x|< n-1$, such that $\barccc\models \exists \bar x \psi(\bar x, a)$
but $\barccc\not\models \exists \bar x \psi(\bar x, c)$. Then $\barccc\models \exists \bar x x' \psi(\bar x,x')\wedge R(x',b)$ but 
 $\barccc\not\models \psi(\bar x,x')\wedge R(x',d)$. But this would mean that $b\not\equiv_n d$. Notice that 
the assumption that $\ccc$ is a VTDAG was used here.\eop

{\bf Proof of Lemma \ref{normali}}
Suppose $\Psi$ is a color closed counterexample of the form ($\heartsuit$). 
Consider the color of $z$ (call it color($z$)). 
More precisely, color($z$) is the color that is enforced by  $\Psi$  on any valuation of  $z$ that satisfies $\Psi$ . 
What we are interested in is not really the full information about  color($z$), but its lightness -- the information 
about the  isomorphic type of ${\cal P}(e)$ for any $e\in \barccc$ having the color that $\Psi$ enforces on $z$.

 Now please be ready for the most complicated argument of this paper. 
Let $e$ be like in the previous paragraph. The set
${\cal P}(e)$ contains some elements $e'$ and $e''$ 
such that $R_1(e',e) \wedge R_2(e'',e)$ are true in $\barccc$. It follows from Definition
 \ref{vtdag} 
that in such  case there must be an atom $Q(e',e'')$  true in  $\barccc$, where  
$Q(e',e'')$ is either $P(e',e'')$ or $P(e'',e')$ for some relation $P\in \Sigma$, or $e'=e''$ (this happens when $R_1=R_2$).
Notice that the atom $Q$ only depends on the color of $z$ not on the choice of $e$. Suppose $Q$ is $P(e',e'')$, the 
other two possibilities are analogous. Now, we claim that  
$\Phi = \psi(\bar x,y)\wedge R_2(z'',z)\wedge P(z',z'')$ is also a counterexample,
with $\sigma_n^\Phi= q_n\circ\sigma_{n+1}^\Psi $,   $e_n^\Phi=e_{n+1}^\Psi$ and $\sigma_n$. 

Notice that we use the notation $q_n$ here in the sense defined in \rpiki{naduzycie}: 
$\sigma_{n+1}^\Psi $, for an argument being a variable of $\Psi$ (or $\Phi$ -- they have the same set of variables) returns 
an element of $M_{n+1}(\barccc)$ and $q_n$ for an argument from $M_{n+1}(\barccc)$ returns an element of $M_{n}(\barccc)$.

We need to show that the conditions from $(\clubsuit)$ are now satisfied. 
It is easy to see that $\sigma_n^\Phi(y)= 
q_n\circ\sigma_{n+1}^\Psi(y) =
q_n(e_{n+1}^\Psi)=
q_n(e_{n}^\Phi)$.

What remains to be shown is that for each $n\in\mathbb{N}$:

\hspace{5mm} (*)  $M_n(\barccc)\models \sigma_n^{\Phi}(\Phi)$~~
and ~~ (**) $\barccc\not\models \exists \bar x \Phi(\bar x, e_n^\Phi)$. 

Let us begin with (**), which is easier. Suppose
$\barccc\models \Phi(\bar x, e_n^\Phi)$. So there exists a valuation $\gamma:Var(\Phi)\rightarrow \barccc $,
with $\gamma(y)=e_n^\Phi$,
such that $\barccc\models\gamma(\Phi)$. Notice that $\gamma(y)=e_{n+1}^\Psi$. We claim that $\barccc\models\gamma(\Psi)$
and this will be  in contradiction with what we assumed about $\Psi$ and $e_{n+1}^\Psi$.  

For the proof of the last claim it is enough to show that $\barccc\models R_1(\gamma(z'),\gamma(z))$, as this is 
the only atom of $\Psi$  missing in $\Phi$. But this follows from
what we know about the isomorphic type of $\gamma(z)$, from the fact that 
 $\barccc\models P(\gamma(z'),\gamma(z'')) \wedge R_2(\gamma(z''),\gamma(z))$ and from the assumption that
the in-degree of each of the relations in $\barccc$ is at most 1 (first condition in Definition \ref{vtdag}). 

Now we are going to prove (*). We know that $M_{n+1}(\barccc)\models \sigma_{n+1}^{\Psi}(\Psi)$, so 
also $M_{n}(\barccc)\models q_n\circ\sigma_{n+1}^{\Psi}(\Psi)$. What remains to be proved is that

(*) $M_{n}(\barccc)\models P(q_n\sigma_{n+1}^{\Psi}(z'),q_n\sigma_{n+1}^{\Psi}(z'')).$

We know that
$M_{n}(\barccc)\models R_1(\sigma_{n+1}^{\Psi}(z'),\sigma_{n+1}^{\Psi}(z))$ and that
$M_{n}(\barccc)\models R_2(\sigma_{n+1}^{\Psi}(z''),\sigma_{n+1}^{\Psi}(z))$. This means that
there are elements $a', a, b'',b$ of $\barccc$ such that:
$q_{n+1}(a)=q_{n+1}(b)=\sigma_{n+1}^{\Psi}(z)$, ~$q_{n+1}(a')=\sigma_{n+1}^{\Psi}(z')$,  ~$q_{n+1}(a'')=\sigma_{n+1}^{\Psi}(z'')$
and $\barccc\models R_1(a',a)\wedge R_2(b'',b)$. The color of $a$ and of $b$ is the color of $z$, so the isomorphic type of
${\cal P}(a)$ is the same as the isomorphic type of
${\cal P}(b)$, and the same  as the isomorphic type of  ${\cal P}(e)$, where $e$ is as in the beginning of Lemma \ref{normali}.
This means that there is an element $a''\in \barccc$ such that $\barccc \models R_2(a'',a)$ and $\barccc \models P(a'',a')$. 
There is no reason to think
that $a''\equiv_{n+1} b''$. But from Lemma \ref{wygodny} we get that
 $a''\equiv_n  b''$. So $a''$ and $a$ are two elements of $\barccc$
such that $\barccc \models R_2(a'',a)$, that $q_n(a)=q_n\sigma_{n+1}^{\Psi}(z)$ and that $q_n(a'')=q_n\sigma_{n+1}^{\Psi}(z'')$.\eop

This ends the  proofs of  Lemmas \ref{niesk-cykl},\ref{normali} , \ref{klaim} and  \ref{duzy}.


\section{Discussion}\label{dyskusja}

\subsection{Beyond the binary case (slightly)}\label{ubranie}

As a careful reader might already have noticed, our proof of Theorem \ref{glowne} can also be read as
a proof of:

\begin{theorem}\label{niebinarne}
Let $\cal T$ be  a set of existential TGDs and plain datalog rules, with each of its existential TGDs of the form:
$\Psi (\bar x, y) \Rightarrow \exists \bar z\; \Phi(y, \bar z)$.
Then, if $\cal T$ is BDD, then it is also FC. 
\end{theorem}

It is because in the proof of Theorem \ref{glowne} we only used the binarity assumption for heads of existential TGDs.  


Notice that we can rewrite existential TGDs from Theorem \ref{niebinarne} into conjunction of existential TGDs with binary heads and some arbitrary datalog rules.
Hence the whole proof of Theorem  \ref{glowne} survives.

Hint: For each  TGD $\Psi (\bar x, y) \Rightarrow \exists \bar z\; \Phi(y, \bar z)$ 
we add new relational symbols $R^1_{\Phi}(y,z_1)\ldots R^n_{\Phi}(y,z_n)$ where $n=|\bar z|$.
We add to the theory rules $\Psi (\bar x, y) \Rightarrow \exists \bar z\; R^i_\Phi(y, z_i)$
and datalog rules $R^1_{\Phi}(y,z_1)\wedge \ldots  \wedge R^n_{\Phi}(y,z_n) \rightarrow  \Phi(y, \bar z)$.

\subsection{The ternary case}

Usually, once we know that some property holds for binary signatures, it is easy to prove, by some sort
of reduction, that it remains true in  the general case. This rule does not seem to be valid
for the BDD/FC conjecture. What we can however easily show is:

\begin{theorem}
If the BDD/FC conjecture for ternary signatures is true then it is  true in  the general case.
\end{theorem}

Instead of presenting a detailed proof of the theorem, which would be boring, let us show  an example of how the reduction 
works. Suppose we have a theory $\cal T$ with a rule like:

(*) $P(x,y,z,x)\Rightarrow \exists t\; R(x,y,z,t) $

then  rewrite it into the following three rules:

 $P(x,y,z,x)\Rightarrow \exists w_1\; R_1(x,y,w_1) $

$P(x,y,z,x) \wedge R_1(x,y,w_1) \Rightarrow \exists w_2\; R_2(w_1,z,w_2) $

$P(x,y,z,x) \wedge R_1(x,y,r)\wedge R_2(r,z,s) \Rightarrow \exists t\; R'(s, t) $

The idea is here that using ternary predicates we can give names to lists of variables, in the good old Prolog way.
We appear to still have non-ternary predicates in the bodies of the rules. But just don't think of them as of predicates any more!
The $P$ in the body of (*) is just a view over the real predicates $P_1$, $P_2$ and $P'$ now, 
which relate to $P$ in the same was as $R_1$, $R_2$ and $R'$
relate to $R$. 

In this way we constructed a new, ternary  theory, call it ${\cal T}'$. What we would now need to show (if it was a real detailed 
proof)  would be that (i) if $\cal T$ is BDD then ${\cal T}'$  also is, and (ii) if ${\cal T}'$ is FC then  $\cal T$ also is. 
To see how (ii) works take a database instance $D$ and query $Q$. Rewrite $D$ and $Q$ into $D'$ and $Q'$ in 
the new ternary language 
(possibly adding some new elements to denote lists of elements of $D$). Of course if $ Chase({\cal T},D)\not\models Q$ then also
 $ Chase(D',{\cal T}')\not\models Q'$. So, if ${\cal T}'$ is FC, 
 there exists a finite ${\cal M}'$ being  a model of   ${\cal T}'$ and $D'$ such that
 ${\cal M}'\not\models Q'$. Now, to finish the proof of (i), define the relations of $\cal M$ as views over respective relations in  ${\cal M}'$.

Showing (i) is not really hard either.

\subsection{Multi-head TGDs}\label{multihead}

The TGDs we consider in this paper are assumed to be single-head. 
Of course if the arity is not restricted, then the validity of the 
BDD/FC conjecture does not depend on this assumption, as every multi-head TGD $\Psi$ can be replaced by a single-head TGD
having, as its head, the join of all the atoms in the head of $\Psi$, and by some datalog rules splitting this join 
back into smaller atoms. But such a simple transformation is not possible for binary signatures. 
It is actually easy to see that the BDD/FC conjecture for multi-head TGDs over binary signatures is already equivalent to
the full conjecture, as any ternary  Datalog$^\exists$ program can be encoded in this format. For example the rule:

$P_1(x,y,z)\wedge P_2(x,y,z')\Rightarrow \exists w\; P(x,z,w)$
 
can be encoded as (read $A^i(t,x)$ as ''$x$ is the $i$'th argument in the atom $t$''):

$A_{P_1}^1(t_1, x)\wedge A_{P_1}^2(t_1, y)\wedge A_{P_1}^3(t_1, z)\wedge$
$A_{P_2}^1(t_2, x)\wedge A_{P_2}^2(t_2, y)\wedge A_{P_2}^3(t_2, z') \Rightarrow $
$\exists t \; A_{P}^1(t, x)\wedge A_{P}^2(t, y)$ 

and $A_{P}^1(t, x)\wedge A_{P}^2(t, y) \Rightarrow \exists w \; A_{P}^3(t, w)$.

\subsection{Why $M_n$ are too poor to be models\\ (in the non-binary case)}\label{why}

The main idea of our proof of Theorem \ref{glownewintro} was first to 
find, for a BDD theory $\cal T$ and a database instance $D$ the skeleton $\ccs$ which is 
 a substructure of $Chase(D, {\cal T})$
 on one hand being simple enough to 
be ptp-conservative, but on the other hand not only 
containing  all the elements of $Chase(D, {\cal T})$, but also  sufficient information
about the relations between elements which require a witness and the witnesses. 
Then the idea was to prove (Lemma \ref{wystarczy}) that the finite model $M_n$ constructed from this simple structure by a quotient operation
can be saturated, using the datalog rules from $\cal T$, to a model of $\cal T$, without adding any new elements being necessary.

The first reason this line of reasoning cannot be used in the general (non-binary) case is that the distinction between
existential TGDs and plain datalog rules makes then no sense any more: each datalog rule can be turned into a TGD by adding a new 
(existentially quantified) dummy variable to the atom on the right hand side of the query. But what we view as
an even  more serious obstacle is that, as the following example shows,
 it is hard to imagine how anything analogous to Lemma  \ref{wystarczy} could be true in the general case:
  
Let the rules of $\cal T$ be: \hfill $R(x,x',y,z) \Rightarrow E(y,z)$

and \hfill $E(x,y),E(t,y)\Rightarrow \exists z\; R(x,t,y,z)$

 and let a database instance $D$ be $\{E(a,b)\}$.

Clearly, $\cal T$ is BDD. And $\ccc = Chase(D, {\cal T})$ is a very simple structure: an infinite $E$-chain,
with additional atom $R(x,x,y,z)$ for each three consecutive elements $x$, $y$, $z$ of this chain. But whenever
any two elements of $\ccc$ are identified by a quotient operation, a new tuple 
satisfying the body of the (only) TGD form $\cal T$ emerges (something we have already seen 
in Example \ref{jakniemozna}), and a new witness $z$ is required for this tuple. Since  the new witness
is a function of the whole tuple, not just of (the element substituted for) $y$, the (already existing) element $t$ of $M_n(\barccc)$
such that $E(y,t)$ cannot be used now, and a new one must be created. If there was just one element this would be something 
we could live with -- our main goal is just to keep the structure finite. But notice that once the new witness $z$, with 
$E(y,z)$ is created, it enforces a new infinite $E$-chain to be built.

\subsection{Beyond BDD. The dead end of the ordering conjecture.}\label{porzadki}

Anyone asked to give some  examples of  theories which are not FC will 
begin from the most natural one --  the infinite total ordering from Remark \ref{teoriazporzadkiem}.

And  it is not immediately clear  how to come out with something really different. For quite some time, we 
believed that the following conjecture could be true: 

\begin{conjecture}[False]\label{false}
$\cal T$ is not FC if and only if $\cal T$ defines an ordering, by which we mean that 
there exists a database instance $D$, an infinite set $A\subseteq Chase(D, {\cal T})$
and a query $\Phi(x,y)$, which is a conjunctive query  with projections, with two free variables,
such that  $ Chase(D, {\cal T})\not\models\exists \bar x \Phi(x,x) $ and  $\Phi$ defines a strict total ordering on $A$.
\end{conjecture}

Notice how beautiful it would be. Even if our BDD/FC conjecture is true (which we believe it is) it does not give
a full explanation of the phenomenon of Finite Controllability, as they are many theories (for example guarded) which 
are FC but not BDD.  Had Conjecture \ref{false} be true, it would have given a sort of such explanation, and a very elegant one,
since the above property of ''defining an ordering'' is very close to (the negation of) the standard, and very important, 
model-theoretic
notion of stability \cite{S69}. Besides, it could give the BDD/FC conjecture as a corollary, if we only could 
prove that a BDD theory never defines an ordering, which we believe should not be very hard.

Clearly, the ''if'' implication  of the conjecture holds true: if $D$, $A$ and $\Phi$ like in the conjecture existed, then
$\exists x\Phi(x,x)$ would be a query  false in  $Chase(D, {\cal T})$ but true in each finite model of ${\cal T},D $ (as each such finite
model must contain a homomorphic image of  $Chase(D, {\cal T})$, and some two elements of $A$ must be mapped, by this 
homomorphism,  to the same 
element of the finite structure)

However, as the following notorious example shows, the opposite implication is not true.
Let $\cal T$ be:

$E(x,y)\Rightarrow \exists z \; E(y,z)$

$R(x,y),E(x,x'),E(y,z),E(z,y')\Rightarrow R(x',y')$

It is not hard to see that $\cal T$ does not define an ordering. We are going to show that $\cal T$ is not FC.
Let  $D$ consist of the atoms $E(a_0,a_1)$ and $R(a_0,a_0)$. 

Then $\ccc=Chase(D, {\cal T})$ is an infinite 
$E$-chain like in Example \ref{oczywisty}, but with additional atoms $R(a_i,a_{2i})$ for each $i$. 
Let $\Phi(x,y)=E(x,y)\wedge R(y,y)$. Clearly, $\ccc\not\models \Phi$ -- the only element $a_0$ of $\ccc$ which
satisfies $R(y,y)$ has no $E$-predecessor. But, as we are going to prove,
if $\cal M$ is any finite model of ${\cal T},D$ then ${\cal M}\models \Phi$.

Indeed, whatever the structure $\cal M$ is, it must contain a sequence $a_0,a_1,\ldots a_m \ldots a_{m+n}$ of elements such that
$a_m = a_{m+n}$ and ${\cal M}\models E(a_i,a_{i+1})$ for each $i<m+n$. 
The datalog rule form $\cal T$ can then prove that ${\cal M}\models R(a_m,a_{m+(m\mod n)})$, and then 
that ${\cal M}\models R(a_{m+l \mod n},a_{m+(m+2l\mod n)})$. Let  $l=-m \mod n$.  Now take $y=a_{m+l \mod n}$ and
$x= a_{m+l-1 \mod n}$ to get $R(y,y)$. 

It is worth mentioning that the structure $\ccc$  from the above example is  ptp-conservative. 
As the degree of the elements of $\ccc $ is bounded by 4, this follows from:

\begin{lemma}
Each binary structure of bounded degree is ptp-conservative. 
\end{lemma}

{\em Proof (hint):} For a given number $m$, color the structure in such a way, 
that  each neighborhood of radius $m$ consists of elements whose colors are pairwise different. 
Then mimic the reasoning from Section \ref{odlozona}.\eop

This shows  ptp-conservativity of $Chase$, which -- as explained in Remark \ref{goodfor} -- guarantees that, if only $n$ is big enough,
 $M_n(\overline{Chase})$ will be  a model
for all the existential TGDs from the theory, does not buy us much  more: the devil can very well be in the plain datalog rules. Notice however,
that not all datalog rules are troublemakers:

\begin{remark}\label{monadycznedatalogi} Suppose $\barccc $ is $n$-conservative up to the size $m$.
 Let $\Psi \Rightarrow Q(x)$ be a datalog rule with at most $m$ variables and 
 with a unary predicate in the head.
 If this true in $\barccc$ then it is also true in $M_n(\barccc)$. Proof: positive $m$-types of $x$ and of $q_n(x)$ are the same.
\end{remark}

\subsection{Guarded TGDs}\label{guarded}

Guarded Datalog$^\exists$ programs, proved to be FC in \cite{BGO10}, consist of guarded rules (datalog rules and TGDs) which 
have  an atom in the body, called the guard,  containing all the variables that occur in the body of this rule. There 
is no restriction on the arity, in particular on the arity of the predicates in the heads of TGDs.

The witness generated by a guarded TGD appears to depend on all the variables in the head of the rule, and in consequence
such a rule seems to be inherently non-binary, not even in the broad sense of Section \ref{ubranie}. 
But, as it turns out, Guarded Datalog$^\exists$ programs are binary in disguise. And, while they are not BDD, still nothing
beyond the techniques developed in Sections  \ref{teoria} and \ref{odlozona} is needed to prove they are FC. 

To be more precise, suppose there exists a Guarded Program $\cal T$, a database instance $D$ and a query $\Phi$ which are a counterexample for FC. Of course
$D$ can be also hardwired into $\cal T$ so we can assume it is empty.

Now we will show how to rewrite  $\cal T$ and $\Phi$ into a binary signature, without changing their status of a counterexample.
 Then we will use the toolkit from Sections  \ref{teoria} and \ref{odlozona} to show very easily that the resulting binary program is FC.

{\bf (i)} First step is similar to the one in the end of Section \ref{hiding} -- we want the predicates which are in the heads of TGDs 
(the TGPs) to be distinct that
the ones in the heads of datalog rules. We also want the rules to respect the order of variables in atoms -- if $x$ is left of $y$
in some atom in the rule then $x$ never can be right of $y$ in any atom of the same rule.
 This can be done by remembering the order of arguments as 
a part of the name of each predicate. Of course $\Phi$ must be rewritten -- each new predicate is now a disjunction of the old predicates. 
Notice that guardedness implies that if $\cal T$ respects the order of variables, if  
$Chase({\cal T})\models R(\bar a, c)$ for some TGP $R$ and if $Chase({\cal T})\models P(\bar b, c)$ 
then $\bar b\subseteq \bar a$. The elements in $\bar a$ are "parents of $c$", who were present in the atom $R$ when $c$ was born, 
and no rule can add anything else left of $c$ in any atom. 

Rename the variables in each rule in such a way, that the rightmost variable of the guard of each rule is $y$. Call this $y$
the leading variable of the rule. 

{\bf (ii)}
We want the elements to know their parents by name. If $R(x_1,\ldots x_k,y)$ is a TGP in $\cal T$ we add to $T$ new rules:\vspace{2mm}
\\
\hspace*{15mm} $R(x_1,\ldots x_k,y)\Rightarrow F_i(x_i,y)$ \vspace{2mm}
\\
for each $i\leq k$, where $F_i$ are new binary predicates.

{\bf (iii)}
Replace each TGD of the form $\Psi \Rightarrow \phi $, with the leading variable $y$, with all possible rules of the form:

\piki{gwiazdeczka}{~~~~~~~~ $\Psi \wedge F_{i_1}(x_1,y)\wedge \ldots \wedge F_{i_k}(x_k,y)   \Rightarrow \phi$}

where $x_1,\ldots x_k$ are all the  non-leading  variables in $\Psi$  and  $i_1,\ldots i_k\leq K$,
 where $K$ is  the maximal arity of the predicates in $\cal T$. This changes nothing, as the elements 
to be substituted for  $x_i$ must have been some parents of $y$ anyway. 

{\bf (iv)}
Now, again in the manner of Section \ref{hiding} rewrite the current $\cal T$ and $\Phi$ in such a way, that each TGP only occurs in one
rule head (this can be  easily done for the cost of some renaming datalog rules). 

{\bf (v)} Now perform step {\bf (iii)} for the datalog rules of the current program.

At some point in Section \ref{ubranie} we wrote: {\em All we need in the proof in Section \ref{dowod} 
is that (...)
the witness generated by the rule only depends on one element in the body (the $y$), while the 
additional elements in the body are just needed to make sure that $y$ has a positive type  
which allows it to demand a witness.} Notice that this is exactly the case with our program now: 
all rules are in the form \rpiki{gwiazdeczka} and the elements of $Chase({\cal T})$ that can
possibly be substituted for elements of $\bar x$ in the body of such rule are themselves functions of $y$. 
So $t$ is just a function of $y$, not of all the elements of $\bar x$!  The only reason why non-binary predicates could be necessary
does not exist any more. Let us get rid of them.

Since the original $\cal T$ was guarded, each atom $P(\bar a)$ in  $Chase({\cal T})$ was contained in some 
TGP atom $R(\bar b,c)$. Our idea is that full 
information about $P(\bar a)$ will be remembered, {\bf without materializing $P(\bar a)$}, in a monadic way, by the 
element $c$. It will need to remember which of its parents are involved in each predicate. 
Of course it also needs to remember the links to its parents -- this is why the relations $F_i$ were introduced.

{\bf (vi)} Replace each TGD of the form: $\Psi \Rightarrow \exists z \; R(x_1,\ldots x_k,$ $z)$ 
($x_k$ may, or may not, by equal to $y$) by the following rules:
$\Psi \Rightarrow \exists z \; E^R(y,z)$ and $\Psi \wedge E^R(y,z)\Rightarrow R^m(z)$ where $E^R$ is a new binary predicate.
$E^R(y,z)$ means something like "the (unique) rule which derives $R$ was applied to a tuple led by $y$ and a witness $z$ was created". 
The newly created element $z$ must also learn who its parents are. For each  $i\in\{1,\ldots k\}$,  if $F_j(x_i,y)$ was an atom in $\Psi$, 
add to 
the current $\cal T$ the  rule:\vspace{3mm}

($\diamondsuit$) \hspace{15mm} $F_j(x_i,y) \wedge E^R(y,z)\Rightarrow F_i(x_i,z)$

and replace each TGP atom $R(x_1,\ldots x_k,z)$ in the body of any rule by:\vspace{2mm}\\
\hspace*{22mm}$F_1(x_1,z)\wedge \ldots F_k(x_k,z)\wedge R^m(z)$

Now the program does not have TGPs of arity higher than 2 any more. Notice that for each variable $x\neq y$ in any rule,
there exists $i$ such that the atom $F_i(x,y)$ is in the body of this rule. We are ready to get rid also of the non-TGPs:

{\bf (vii)} In each rule, with the atoms 
$F_{i_1}(w_1,y), \ldots F_{i_1}(w_l,y)$ in its body, replace each occurrence of a non-TGP atom $Q(w_1,\ldots w_l)$ with 
 $Q_{i_1i_2\ldots i_l }(y)$,
where $Q_{i_1i_2\ldots i_l }(y)$ is a new monadic predicate (in which $y$ remembers what his
parents with numbers $ i_1,i_2,\ldots i_l$ are involved in). For each two monadic predicates 
of the form $Q_{i_1i_2\ldots i_l }(y)$ and $Q_{j_1j_2\ldots j_l}$ add to  the program all possible rules
of the form: 

$F_{i_1}(x_1,y)\ldots \wedge F_{i_l}(x_l,y) \wedge F_{j_1}(x_1,z)\ldots \wedge F_{j_l}(x_l,z)\wedge$\vspace{2mm}\\
\hspace*{30mm} $\wedge  Q_{i_1i_2\ldots i_l }(y) \Rightarrow Q_{j_1j_2\ldots j_l}(z)$

The role of the last rule is to make sure that, once an atom of the predicate $Q$, involving $x_1\ldots $ $x_l$ is derived, all the elements that have $x_1\ldots $ $x_l$ among their parents are aware of that and ready to use this fact in further derivations.
 
We now have a new program over a binary signature, 
call it ${\cal T}'$. It follows from the construction that $\ccc = Chase({\cal T}')$ is almost the same structure as $Chase({\cal T})$ (where $\cal T$ is the original guarded program). They both have the same elements, and the predicates of each of them can be seen as views over the predicates of the other one.
But notice that $Chase({\cal T}')$ is a binary structure satisfying the assumptions of of Lemma \ref{duzy}.  So it is ptp-conservative. This means that if $n$ is big enough then  $M_n(\barccc)$ is a model 
for all the existential TGDs in ${\cal T}'$ and that $M_n(\barccc)\not\models \Phi'$ (where $\Phi'$ is the original query $\Phi$ after all the rewritings). To finish the proof of FC for Guarded Datalog$^\exists$ programs we only need to show that $M_n(\barccc)$ is also a model
of all the datalog rules in ${\cal T}'$. All the datalog rules except from the rules of the form ($\diamondsuit$) have a unary
atom in the head, so (by Remark \ref{monadycznedatalogi}) we do not need to bother about them at all. What remains to be seen 
is that the rules of the form ($\diamondsuit$) also remain true in $M_n(\barccc)$. So suppose 
$M_n(\barccc)\models F_j(a,b) \wedge E^R(b,c)$ for some $a,b,c$. This means that there exist $a',b',b'',c''$ in $\barccc$ such that
$\barccc \models F_j(a',b')$,  $\barccc \models E^R(b'',c'')$, $q_n(a')=a$, $q_n(b')=q_n(b'')=b$ and $q_n(c')=c$.
But $\barccc$ can be seen as a Chase of the guarded theory $\cal T$ with the natural coloring, so the types of successors of 
an element only depend on the  type of this element, and it is easy to see that if $b'\equiv_n b''$ and
 $\barccc \models E^R(b'',c'')$ then there must exist  $c'\equiv_n c''$ such that  $\barccc \models E^R(b',c')$. Since
the rule ($\diamondsuit$) was true in $\barccc$ we get that $\barccc\models F_i(a',c')$ which implies that 
$M_n(\barccc)\models F_i(a,c)$. We proved that $\diamondsuit$ remains true in $M_n(\barccc)$.


\begin{thebibliography}{13}
 

 \bibitem{BGO10} V. Barany, G. Gottlob, and M. Otto. {\em Querying the guarded fragment};  Proc. of the 25th 
IEEE Symposium on Logic in Computer Science, LICS 2010, Edinburgh, UK, pp. 1-10, 2010;

\bibitem{CGT09} A. Cali, G. Gottlob, and T. Lukasiewicz;  {\em A general datalog-based framework for tractable query
answering over ontologies}; in Proc. of PODS, 2009;

\bibitem{CGT12} A. Cali, G. Gottlob, and T. Lukasiewicz;  {\em A general datalog-based framework for tractable query
answering over ontologies}; J. Web Sem. 14, 2012, 57-83 

\bibitem{CGP10}  A. Cali, G. Gottlob, and A. Pieris; {\em Advanced processing for ontological queries}; Proc. VLDB-10,
3(1):554-565, 2010; 

\bibitem{CGP10'}  A. Cali, G. Gottlob, and A. Pieris;  
{\em Query Answering under Non-guarded Rules in Datalog+/- }; Web Reasoning and Rule Systems
Lecture Notes in Computer Science, 2010, Volume 6333, pp 1-17;

\bibitem{GM13} T. Gogacz, J. Marcinkowski; ~~{\em Converging to the Chase and Some Finite. Controllability Results}; 
Proc. of the 28th IEEE Symposium on Logic in Computer Science, LICS 2013, New Orleans, USA, to appear;

\bibitem{JK84} D. S. Johnson and A. C. Klug. {\em Testing containment of conjunctive queries under functional and inclusion
dependencies}; JCSS 28(1):167-189, 1984;

\bibitem{R06} R. Rosati; {\em On the decidability and finite controllability of query processing in databases with incomplete
 information}; in Proc. PODS 2006, pp. 356--365;

\bibitem{R11} R. Rosati; {\em On the decidability and finite controllability of query processing in databases with incomplete
 information}; J. Comput. Syst. Sci. 77(3),2011,  pp. 572-594

\bibitem{S69} S. Shelah; (1969), {\em Stable theories}; Israel J. Math. 7 (3): 187-202.

\end{thebibliography}
\end{document}